\documentclass[10pt,pra,aps,twocolumn,showpacs,notitlepage,nofootinbib,superscriptaddress]{revtex4-1}

\usepackage [english]{babel}
\usepackage [utf8]{inputenc}

\usepackage{graphicx}
\usepackage{amsmath}
\usepackage{amssymb}
\usepackage{ulem}
\usepackage{times}
\usepackage{color}
\usepackage{bm}
\usepackage{xcolor}
\usepackage{natbib}
\usepackage{subfigure}
\usepackage[colorinlistoftodos, shadow,textsize= footnotesize]{todonotes}
\usepackage[linktocpage=true,colorlinks]{hyperref}

\newcommand{\vect}[1]{\mathbf{#1}}

\newcommand{\ket}[1]{|#1\rangle}
\newcommand{\bra}[1]{\langle#1|}
\newcommand{\braket}[2]{\langle#1|#2\rangle}
\newcommand{\mean}[1]{\ensuremath{\langle #1 \rangle}}

\newcommand{\rref}[1]{(\ref{#1})}

\newcommand{\be}{\begin{equation}}
\newcommand{\ee}{\end{equation}}

\newcommand{\ben}{\begin{equation*}}
\newcommand{\een}{\end{equation*}}

\newcommand{\ud}{\mathrm{d}}

\newcommand{\beq}{\begin{eqnarray}}
\newcommand{\eeq}{\end{eqnarray}}

\begin{document}

\title{Fast and optimal generation of entanglement in bosonic Josephson junctions}

\author{Giacomo Sorelli}
\affiliation{QSTAR, INO-CNR and LENS, Largo Enrico Fermi 2, I-50125 Firenze, Italy}
\affiliation{Physikalisches Institut, Albert-Ludwigs-Universit\"at Freiburg, Hermann-Herder-Stra\ss e 3, D-79104 Freiburg, Germany}

\author{Manuel Gessner}
\affiliation{QSTAR, INO-CNR and LENS, Largo Enrico Fermi 2, I-50125 Firenze, Italy}

\author{Augusto Smerzi}
\affiliation{QSTAR, INO-CNR and LENS, Largo Enrico Fermi 2, I-50125 Firenze, Italy}

\author{Luca Pezz\`e}
\affiliation{QSTAR, INO-CNR and LENS, Largo Enrico Fermi 2, I-50125 Firenze, Italy}

\begin{abstract}
We use an exact quantum phase model to study the dynamical generation of particle-entanglement in a bosonic Josephson junction composed
by two coupled and interacting Bose-Einstein condensates.
Using analytical arguments, we show that linear coupling can accelerate the creation of entanglement with respect to the well known one-axis twisting model where coupling is absent.
Furthermore, with a numerical analysis of optimal control schemes, we identify the optimal parameters for the fast generation of metrologically useful entanglement on short time scales: the optimal evolution is generated for a specific ratio between the coupling and the interaction strength.
\end{abstract}

\date{\today}


\maketitle

\section{Introduction}
The generation of multiparticle entanglement in ultracold atomic gases is a vivid area of research.
Besides the foundational interest, multiparticle entanglement can find important applications in quantum metrology~\cite{PezzeRMP}.
For this purpose, it is desirable to develop protocols creating entanglement on the fastest possible time scales.
One of the most successful protocols for the dynamical creation of multiparticle entanglement 
is one-axis twisting  -- originally proposed by Kitagawa and Ueda~\cite{KitagawaPRA1993} for spin-1/2 particles.
For ultracold atomic gases, it corresponds to the unitary evolution $e^{-i \hat{H}_{\rm OAT} t/\hbar}$ generated by the quadratic Hamiltonian
$\hat{H}_{\rm OAT} = \hbar \chi \hat{J}_z^2$, which expresses the action of elastic particle-particle collisions of strength $\hbar \chi$~\cite{SorensenNATURE2001}.
Here, 
$\hat{J}_x = (\hat{a}^\dagger\hat{b}+ \hat{b}^\dagger \hat{a})/2$, 
$\hat{J}_y = (\hat{a}^\dagger\hat{b}- \hat{b}^\dagger \hat{a})/(2i)$, 
and 
$\hat{J}_z = (\hat{a}^\dagger\hat{a}- \hat{b}^\dagger \hat{b})/2$ 
are pseudo-spin operators satisfying the commutation relation $[\hat{J}_l, \hat{J}_n] = i \sum_m \epsilon_{lnm}\hat{J}_m$, where 
$\epsilon_{lnm}$ is the Levi-Civita symbol, and  $\hat{a}$ ($\hat{a}^\dag$) and $\hat{b}$ ($\hat{b}^\dag$) are bosonic mode annihilation (creation) operators.
The generation of entanglement via one-axis twisting dynamics has been 
realized experimentally with Bose-Einstein condensates~\cite{GrossNATURE2010, RiedelNATURE2010, OckeloenPRL2013, MuesselPRL2014, SchmiedSCIENCE2016} 
and cold thermal atoms~\cite{LerouxPRL2010}.
In the latter case, off-resonant atom-light coupling in an optical cavity is used to realize 
an effective atom-atom interaction~\cite{TakeuchiPRL2005, Schleier-SmithPRA2010}. 
The physics of the one-axis twisting model is enriched when the quadratic Hamiltonian, responsible for the twisting,
competes with a linear coupling terms between the two modes that turns the state along an orthogonal direction 
\be \label{H}
\hat{H} = \hbar\chi  \hat{J}_z^2 - \hbar \Omega \hat{J}_x.
\ee
This Hamiltonian describes a bosonic Josephson junction (BJJ) made of two weakly-coupled and 
interacting Bose-Einstein condensates in the presence of an external field 
of strength $\hbar\Omega$~\cite{MilburnPRA1997, CiracPRA1998, SteelPRA1998, AnanikianPRA2006, GatiJPA2007}.
The model \rref{H} can be realized in a physical system employing either external 
(e.g. a double-well trapping potential) or internal (e.g. two hyperfine atomic states) 
degrees of freedom of an atomic Bose-Einstein condensate~\cite{PezzeRMP}.
The dynamics generated by $\hat{H}$ is particularly interesting
when the coherent spin state is prepared in correspondence to the mean-field unstable fixed 
point~\cite{SmerziPRL1997,RaghavanPRA1999,MicheliPRA2003,StrobelSCIENCE2014}.
The dynamical evolution within the model (\ref{H}) has also been explored for the generation of spin-squeezed states
via shortcuts to adiabaticity~\cite{JuliaDiazPRA2012b} and catlike states by optimal control methods~\cite{LapertPRA2012}.
In recent experiments~\cite{MusselPRA2015} it has been shown that 
the twist-and-turn dynamics generates entanglement on a rate faster than the one-axis twisting dynamics, even for short times. 

Here we study the BJJ dynamics within the exact quantum phase model (EQPM) discussed in Ref.~\cite{AnglinPRA2001}, 
see also \cite{PezzePRA2005}, 
and use the quantum Fisher information and the spin-squeezing parameter as entanglement witness. 
Our analysis extends previous results for the BJJ dynamics studied in 
the weak interaction limit \cite{LawPRA2001} or short time scales \cite{JuliaDiazPRA2012}.
We show that, for sufficiently long times, there is a substantial difference with the one-axis twisting dynamics, 
and provide optimal values of the ratio $\Lambda = N \chi/\Omega$ for fast creation of entanglement.
Our results provide an analytical understanding of experimental observations \cite{MusselPRA2015, StrobelSCIENCE2014}.
Finally, we provide numerical evidence for the optimality of the working point $\Lambda=2$ for fast entanglement generation on relatively short time scales 
by comparing to arbitrary dynamical processes of the bosonic Josephson junction.
The approximations used to solve the EQPM dynamics fail in the long-time limit where 
the system shows the onset of a Schr\"odinger cat state~\cite{MicheliPRA2003}. 

The manuscript is organized as follows. 
We first introduce, in Sec.~{\ref{Sec2}}, the EQPM and the relevant multi-particle entanglement witnesses.
In section \ref{secpi} we present the dynamical generation of entanglement when the initial state 
is a coherent spin state pointing in the negative direction of the $x$-axis of the Bloch sphere.
This situation corresponds to the unstable regime of the mean field dynamics
and it is the working condition of recent experiments \cite{MusselPRA2015, StrobelSCIENCE2014}.
In Sec.~\ref{secOAT} we draw a direct comparison with the one-axis twisting dynamics. 
In Sec.~\ref{Sta} we compare with the analysis of the dynamical generation of entanglement 
when the system is prepared in a coherent spin state pointing in the positive $x$-axis of the Bloch sphere, corresponding to the   
the fixed stable point of the mean field dynamics. 
Finally, in Sec.~\ref{Sta} we discuss optimality of the unstable BJJ dynamics. 

\begin{figure}[t!]
\includegraphics[width=0.46\textwidth]{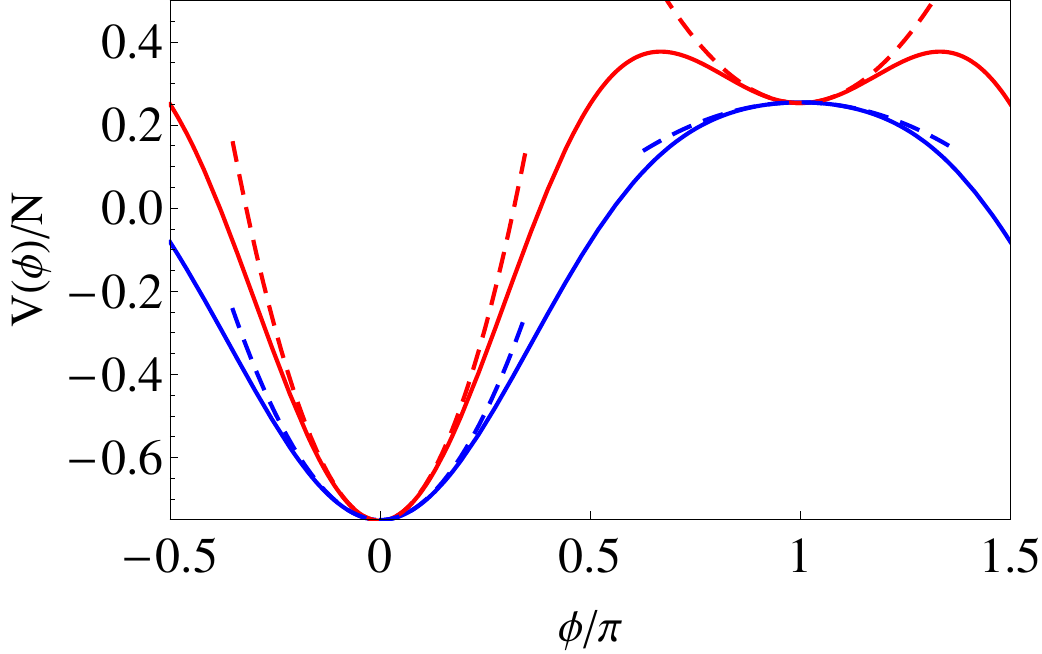}
\caption{(color online) Effective phase potential $V(\phi)$ (solid lines), Eq.~(\ref{VEQPM})
normalized by $N$ in the limit $N \gg 1$.
Different lines refer to different values of $\Lambda$: the red line is for $\Lambda = 0.5$ while the blue one is for $\Lambda = 1.5$.
Dashed lines are the harmonic oscillator approximation in the nearby of $\phi=0$ and $\phi=\pi$. } 
\label{figHarm}
\end{figure}

\section{Two-mode dynamics of interacting bosons}
\label{Sec2}

\subsection{Exact quantum phase model}
\label{EQPMsec}

The Schr\"odinger equation $i \hbar \tfrac{\partial \ket{\psi(t)}}{\partial t} = \hat{H} \ket{\psi(t)}$ describing the 
dynamics of the many-body state $\ket{\psi(t)}$ can be studied within the EQPM~\cite{AnglinPRA2001}.
It consists in expanding a general quantum state of the two-mode system, 
\be
\label{barg}
\ket{\psi(t)} = \int_{-\pi}^{\pi} \frac{\ud \phi}{2\pi} \Psi(\phi,t) \ket{\phi},
\ee
on the overcomplete set of Bargmann states \cite{WallsBOOK} 
\be
\ket{\phi} = \sum_{n=-N/2}^{N/2} \frac{e^{in\phi}}{\sqrt{\left(\frac{N}{2}+n\right)!\left(\frac{N}{2}-n\right)!}} \Big\vert \frac{N}{2}+n \Big\rangle_a \Big\vert \frac{N}{2}-n \Big\rangle_b. \nonumber
\ee
Here $\ket{N/2 + n}_{a}\ket{N/2 - n}_{b}$ indicates a Fock state of $N/2 + n$ particles in mode $a$ and 
$N/2 - n$ particles in mode $b$, where $N$ is the total number of particles.
In this representation we can express the action of spin operators on the state $\ket{\psi}$
in terms of differential operators acting on $\Psi(\phi,t)$:
explicit expressions for the first and second moments of the spin operators are reported in the Appendix \ref{App0}.
Introducing the function $\psi(\phi,t) =  \Psi(\phi,t) e^{-\frac{N}{2 \Lambda} \cos \phi}$,
we can write the Hamiltonian \rref{H} as \cite{AnglinPRA2001}
\be
\label{HEQPM}
\hat{H} \ket{\psi(t)} = \int_{-\pi}^{\pi} \frac{\ud \phi}{2\pi} \left[\mathcal{H}_\phi \psi(\phi,t)\right] 
e^{\tfrac{N}{2\Lambda}\cos \phi}
\ket{\phi}
\ee
where $\Lambda = \frac{N\chi}{\Omega}$ is the ratio between interaction and tunneling. 
In the following we focus on the case of repulsive interaction $\Lambda>0$. 
Our analysis generalizes to attractive interaction $\Lambda<0$ via the mapping $t \to -t$ and $\phi \to \phi+\pi$.
It should be noticed that 
the EQPM~(\ref{HEQPM}) is exact for any number of particles. 
The  Hamiltonian $\mathcal{H}_\phi$ is given by 
\be
\label{EQPM}
\mathcal{H}_\phi = \Omega\left(-\frac{\Lambda}{N}\frac{\partial^2}{\partial \phi^2} +V(\phi) \right).
\ee
where
\be \label{VEQPM}
V(\phi) = - \left(\frac{N+1}{2}\right)\cos \phi - \frac{N}{8\Lambda}\cos 2\phi
\ee
is an effective potential.
The time evolution of the system can thus be mapped on the dynamics of a fictitious wave-packet, 
\be
\label{TE}
i\hbar\frac{\partial \psi(\phi,t)}{\partial t} = \mathcal{H}_\phi \psi(\phi,t),
\ee 
evolving in the potential $V(\phi)$.
In Fig. \ref{figHarm} we show $V(\phi)$ for different values of $\Lambda$.
While the potential is always confining in the vicinity of $\phi = 0$, close to $\phi=\pi$ the potential $V(\phi)$ becomes a repulsive inverted parabola when $\Lambda >  N/(N+1)$.
The different behavior of the effective potential close to $\phi=0$ and $\pi$ has direct implications in the creation of many-particle entanglement, as discussed below. 
In particular, we study the phase dynamics of a Gaussian wave function $\psi(\phi,t)$ centered in $\phi=0$ or $\pi$.
Details on the use of the Gaussian approximation and on the computation of relevant expectation values can be found in Appendix \ref{AppI}.

As a main drawback of the EQPM, the Bargmann basis is overcomplete, which makes the calculation of expectation values of a generic observables $\hat A$
\be
\bra{\psi} \hat{A} \ket{\psi} =  \int_{-\pi}^{\pi} \frac{\ud \theta}{2\pi} \int_{-\pi}^{\pi} \frac{\ud \phi}{2\pi} \Psi(\theta,t)^* A(\phi,t)  \Psi(\phi,t) \braket{\theta}{\phi},
\ee
and the inner product
\be
\braket{\psi}{\psi'} =  \int_{-\pi}^{\pi} \frac{\ud \theta}{2\pi} \int_{-\pi}^{\pi} \frac{\ud \phi}{2\pi} \Psi(\theta,t)^*  \Psi'(\phi,t) \braket{\theta}{\phi},
\ee
non-local in phase, with the non standard inner product $\braket{\theta}{\phi} = (2^N/N!)\cos^N[(\theta-\phi)/2]$. 
This term is crucial to cut Fourier contributions with frequency smaller than $1/N$. 
In the following, we will use the approximation $\cos^N[(\theta - \phi)/2] \sim \exp[-N(\theta-\phi)^2/8]$.
The factor  $e^{\frac{N}{2\Lambda}\cos\phi}$ in Eq.~\rref{HEQPM} is important to obtain the correct dynamics.
It can be conveniently approximated as $e^{-\frac{N}{4\Lambda}\phi^2+\frac{N}{2\Lambda}}$ in the vicinity of $\phi=0$ and as $e^{\frac{N}{4\Lambda}(\phi-\pi)^2-\frac{N}{2\Lambda}}$ around $\phi=\pi$. Both these approximations are justified by the fact that we are working with Gaussian wave packets that are localized in $\phi$.

In a previous work \cite{PezzePRA2005}, the EQPM has been used to study the ground state of Eq.~(\ref{H}) for repulsive interaction. 
Here we consider the dynamical generation of entanglement starting from 
separable states
\be \label{CSS}
\ket{\vartheta,\varphi}_N = \Big( \hat{a}^\dag \cos \frac{\vartheta}{2} +  \hat{b}^\dag e^{i \varphi} \sin \frac{\vartheta}{2} \Big)^N \ket{{\rm vac}},
\ee
where $\ket{\rm vac}$ is the vacuum.
Eq. (\ref{CSS}) is generally indicated as coherent spin state (CSS) \cite{ArecchiPRA1972} and is given by all $N$ qubits pointing along the same direction 
$\vect{s} = \{ \sin \vartheta \cos \varphi, \sin \vartheta \sin \varphi, \cos \vartheta \}$
(called the mean spin direction) in the Bloch sphere, where $0 \leq \vartheta \leq \pi$ and $-\pi \leq \varphi < \pi$.
The CSS has collective spin mean values $\mean{\hat{J}_{\vect{s}}} = N/2$ and $\mean{\hat{J}_{\perp}} = 0$,
and variances  $( \Delta \hat{J}_{\vect{s}} )^2 = 0$ and $(\Delta \hat{J}_{\perp})^2 = N/4$, where 
 $\perp$ indicates an arbitrary direction orthogonal to the mean spin direction.
 
It is worth recalling that there are alternative approaches to the EQPM to study Eq.~(\ref{H}), besides (numerical) exact diagonalization. 
 A well explored method is a semiclassical approximation in number space \cite{ShchesnovichPRA2008, JavanainenPRA1999,JuliaDiazPRA2012}. 
 

\subsection{Entanglement, spin-squeezing and quantum Fisher information}
\label{SSF}

To witness and quantify the creation of particle entanglement we calculate the spin-squeezing parameter 
\be
\xi^2 = \frac{N(\Delta \hat{J}_{\vect{n}_3})^2}{\langle \hat{J}_{\vect{n}_1} \rangle^2}
\ee
and the quantum Fisher information (QFI)~\cite{HelstromBOOK, BraunsteinPRL1994}, which for pure states is given by
\be
F_Q[\hat{J}_{\vect{n}_2}] = 4 (\Delta \hat{J}_{\vect{n}_2})^2, 
\ee
where $\vect{n}_1$, $\vect{n}_2$ and $\vect{n}_3$ are three mutually orthogonal directions.
The criteria \cite{WinelandPRA1992}
\be
\label{xi}
\xi^2< 1
\ee 
and \cite{PezzePRL2009}
\be \label{QFI}
\zeta^2 = \frac{N}{F_Q[\hat{J}_{\vect{n}_2}]}<1.
\ee 
are sufficient conditions for entanglement between the $N$ particles,
and can be extended to witness $k$-partite entanglement as $\zeta^2<1/k$ (and the same for $\xi^2$)~\cite{HyllusPRA2012}.
They have an operational significance.
Eq. (\ref{xi}) recognizes all entangled spin-squeezed states that can be used to sense a rotation 
$e^{-i \theta \hat{J}_{\vect{n}}}$ with a sensitivity overcoming the shot-noise limit $\Delta \theta = 1/\sqrt{mN}$, where $m$ accounts for the repetition of independent measurements.
Since the highest attainable interferometric phase sensitivity is limited by the quantum
Cram\'er-Rao lower bound, $\Delta \theta_{\rm QCR} = 1/\sqrt{m F_Q[\hat{J}_{\vect{n}_2}]}$ \cite{HelstromBOOK, BraunsteinPRL1994}, Eq.~(\ref{QFI}) detects 
all the entangled states that can be used to overcome the shot-noise limit when sensing the rotation $e^{-i \theta \hat{J}_{\vect{n}_2}}$.
In particular, the inequality 
\be
\zeta^2 \leq \xi^2
\ee
holds \cite{PezzePRL2009} and thus the QFI recognizes the full class of states (including the spin-squeezed ones) containing useful entanglement to overcome the shot-noise limit.
The highest value of the QFI is $F_Q[\hat{J}_{\vect{n}_2}] =N^2$, which corresponds to the so called Heisenberg limit of phase sensitivity, 
$\Delta \theta_{\rm HL} = 1/\sqrt{m}N$, representing the ultimate bound for phase estimation \cite{GiovannettiPRL2006}. 

In the following we compute $\xi^2$ and $\zeta^2$ for initial CSS pointing on the $x$-axis and evolving according to Eq.~(\ref{H}).
These states are characterized by $\langle \hat{J}_{z} \rangle = \langle \hat{J}_{y} \rangle=0$ at all times.
Therefore, the optimal spin-squeezing and QFI are obtained by minimizing (for $\xi$) or to maximizing (for $\zeta$) 
the spin variance on the $y-z$ plane. 
This is done by computing the eigenvalues
 \be
 \label{lpm}
 \lambda_\pm = \frac{1}{2} \left[\gamma_{zz} +\gamma_{yy} \pm \sqrt{\left(\gamma_{zz} -\gamma_{yy} \right)^2 +\left(2\gamma_{yz}\right)^2} \right].
 \ee
(with $\lambda_+>\lambda_-$) of the renormalized covariance matrix
\be
\label{cov}
\gamma_{ij} = \frac{2\langle \lbrace\hat{J}_i, \hat{J}_j\rbrace\rangle}{N}, 
\ee
where $i,j=y,z$, and the curly brackets stand for the anti-commutator between the two angular momenta.
Note that $\gamma_{zz} = 4 (\Delta \hat{J}_z)^2/N$ and $\gamma_{yy} = 4(\Delta \hat{J}_y)^2/N$.
The optimal values of $\xi^2$ and $\zeta^2$ are thus given by 
\be \label{spinsq}
\xi^2_{\rm opt} = \frac{N^2\lambda_-}{4 \langle \hat{J}_{x}\rangle^2}, \quad {\rm and} \quad   \zeta^2_{\rm opt} = \frac{1}{\lambda_+}.
 \ee


\section{Entanglement generation around \texorpdfstring{$\mbox{\boldmath $\phi=\pi$}$}{}}
\label{secpi}

We focus here on the generation of entanglement starting from the CSS
\be \label{CSS2}
\ket{\pi/2, \pi}_N = \bigg( \frac{\hat{a}^\dagger - \hat{b}^\dagger}{\sqrt{2}} \bigg)^N\ket{{\rm vac}}, 
\ee
polarized along the negative $x$-axis (i.e., the eigenstate of $\hat J_x$ with minimum eigenvalue $-N/2$) and following a quench 
of $\Lambda$.
For $\phi \approx \pi$, Eq.~\rref{EQPM} becomes the Hamiltonian of a harmonic oscillator of squared frequency 
\be \label{omegapi}
\frac{\omega_\pi^2}{\Omega^2} = 1 - \Lambda \Big(1+\frac{1}{N} \Big).
\ee
Eq. (\ref{omegapi}) is positive [corresponding to a confining effective potential $V(\phi)$] for $\Lambda < N/(N+1)$ and negative 
[corresponding to a repulsive $V(\phi)$] for $\Lambda > N/(N+1)$, see Fig.~(\ref{figHarm}).
The change of sign of $\omega_\pi^2$ is directly linked to the onset on instability of the corresponding mean-field fixed point \cite{SmerziPRL1997,RaghavanPRA1999}. 
We further define $\omega_\pi =  \sqrt{\vert\omega_\pi^2\vert}$.

The exact representation of the coherent states \rref{CSS2} in the EQPM is a Dirac delta, but for the calculation it is convenient to consider a Gaussian wave function.
We initialize the EQPM dynamics with the Gaussian ground state of Eq.~\rref{EQPM} with an initial interaction parameter $\Lambda_0 \neq 0$.
We then quench $\Lambda$ to a final constant value, and compute the evolution of the wavefunction 
\be \label{GState}
\Psi(\phi,t) = e^{-[a_{\pi}(t)+ib_{\pi}(t)](\phi-\pi)^2}.
\ee
The Schr\"odinger equation (\ref{TE}) provides 
\begin{subequations}
\label{abPi}
\begin{align}
&a_{\pi}(t) = \frac{\frac{N\omega_{\pi}^2}{2\Lambda^2}\alpha_\pi}{\frac{\omega_{\pi}^2}{\Lambda^2}+\alpha_\pi^2 + \left(\frac{\omega_{\pi}^2}{\Lambda^2}-\alpha_\pi^2\right)\cos(2\omega_{\pi} t)}-\frac{N}{4\Lambda},\\
&b_{\pi}(t) = -\frac{N\omega_{\pi}}{4\Lambda}\frac{\left(\frac{\omega_{\pi}^2}{\Lambda^2}-\alpha_\pi^2\right)\sin(2\omega_{\pi} t)}{\frac{\omega_{\pi}^2}{\Lambda^2}+\alpha_\pi^2 + \left(\frac{\omega_{\pi}^2}{\Lambda^2}-\alpha_\pi^2\right)\cos(2\omega_{\pi} t)},
\end{align}
\end{subequations}
with $\alpha_\pi  = \big(\frac{1+\omega_\pi(\Lambda_0)}{\Lambda_0}+\frac{1}{\Lambda}\big)$ and $\omega_\pi(\Lambda_0)$ is the frequency Eq.~(\ref{omegapi}) 
computed with an interaction parameter $\Lambda_0$. 
Finally, in order to compute the dynamics of an initial CSS [Eq.~(\ref{CSS2}), corresponding to $\Lambda_0=0$), 
mean spin moments are calculated taking the limit $\Lambda_0 \to 0$, 
see further discussion in Appendix~\ref{AppI}.
We also consider $N \gg 1$ and neglect terms of the order $1/N$.
We emphasize that, within the approximation (\ref{GState}), we cannot distinguish between the QFI and spin-squeezing: $\zeta^2_{\rm opt} = \xi_{\rm opt}^2$ in our analytical calculations.

\begin{figure}[t!]
  \includegraphics[width=0.45\textwidth]{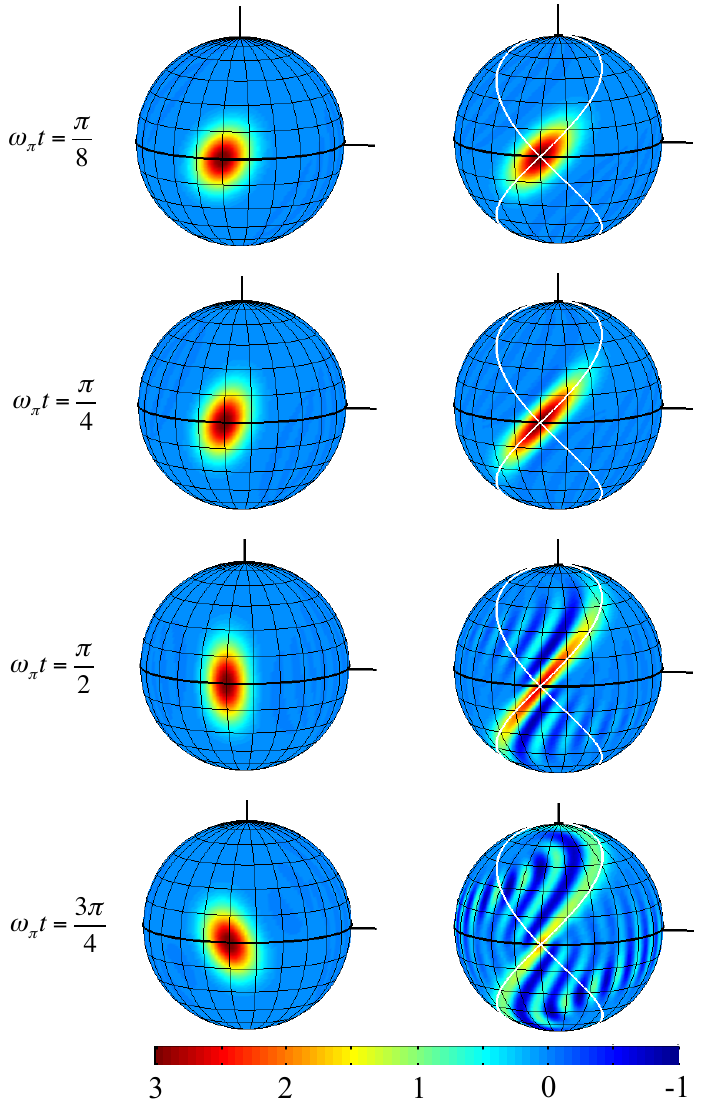}
 \caption{(color online) Wigner distributions (normalized to one) of the states obtained from the dynamical evolution of Eq.~(\ref{CSS2}) in the case $\Lambda=0.5$ (left column) and 
 $\Lambda=2$ (right column). Different rows correspond to different evolution times. 
 The thick white line is the separatrix of the mean-field dynamics. Here $N=30$.}
 \label{HusPi}
\end{figure}

\begin{figure}[t!]
\includegraphics[width=0.46\textwidth]{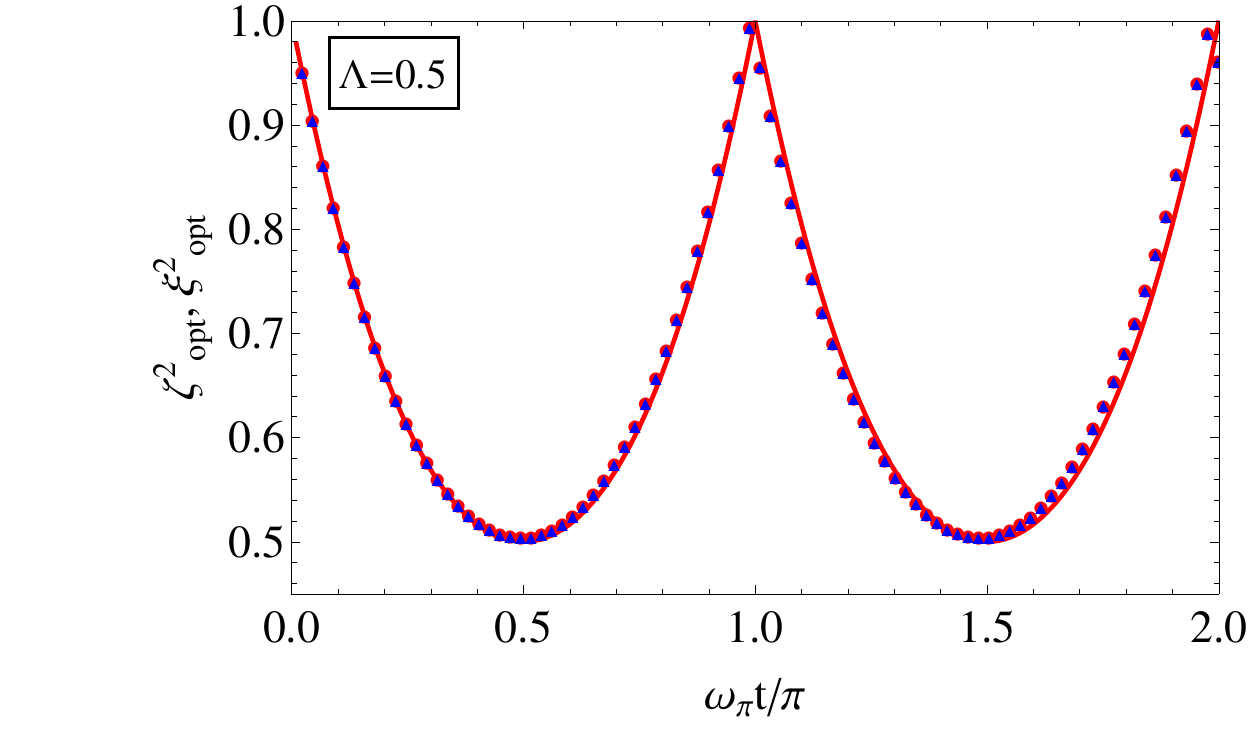} 
\includegraphics[width=0.46\textwidth]{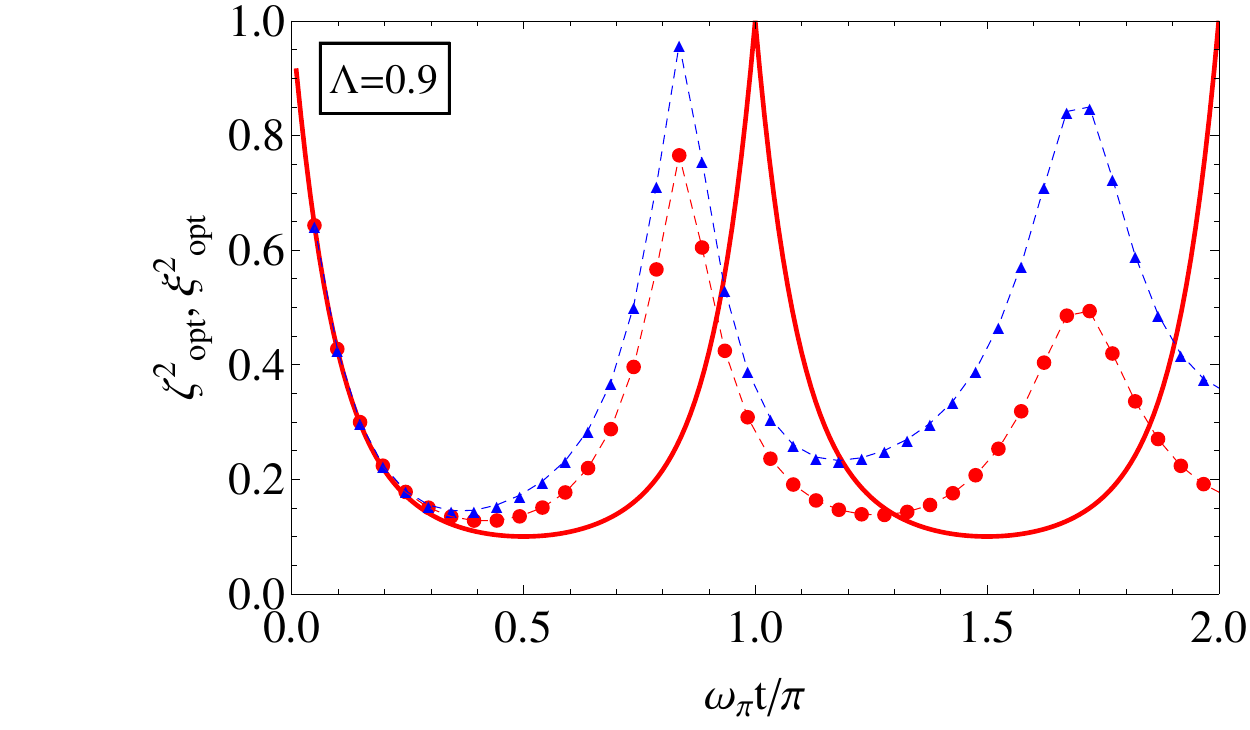}
\caption{(color online) Time evolutions of $\zeta^2_{\rm opt}$ (red points) and $\xi^2_{\rm opt}$ (blue triangles) optimized over 
collective spin directions and computed via exact numerical calculation.
The solid red line is the analytical result for $\zeta^2_{\rm opt}$ in the Gaussian approximation. 
The two panels are obtained for different values of $\Lambda$ and $N=200$, with the initial state Eq.~\rref{CSS2}.} 
\label{figpista}
\end{figure}

{\it Stable regime, $\Lambda < N/(N+1)$.}
In the stable regime, the dynamics is characterized by a periodic generation of phase squeezing.
Fig. \ref{HusPi} (left column) shows the Wigner distribution on the Bloch sphere \cite{DowlingPRA1994}
plotted at different evolution times.
The initial isotropic Wigner distribution of the CSS deforms in time, 
becoming elliptic and reaching its maximum squeezing along the $y$-axis (phase squeezing, see below).
A calculation of Eq.~(\ref{cov}) with the Gaussian state (\ref{GState}) gives (after taking the limit $\Lambda_0\rightarrow 0$)
\begin{subequations}
\label{expvalpista}
 \begin{align}
 &\gamma_{zz}  = \frac{1}{2(\Lambda-1)}\left(\Lambda+\Lambda \cos 2\omega_{\pi} t -2 \right),\\
 &\gamma_{yy} = \frac{1}{2}\left(2-\Lambda+\Lambda \cos 2\omega_{\pi} t \right),\\
 &\gamma_{yz} = \gamma_{zy} = -\frac{2\Lambda}{\sqrt{1-\Lambda}}\sin 2\omega_{\pi} t,
\end{align}
\end{subequations}
and 
\be
\frac{2\langle\hat{J}_x\rangle}{N} = -1 + \frac{1}{4N}\frac{\Lambda^2}{\Lambda-1}\left(\cos 2\omega_{\pi} t-1 \right).
\ee
These expectation values are used to calculate $\zeta^2_{\rm opt}$.
In Fig.~\ref{figpista} we show $\zeta^2_{\rm opt}$ as a function of time and we compare them with the numerical results obtained from the exact diagonalization of Eq.~(\ref{H}).
The upper panel is obtained for $\Lambda=0.5$ and shows periodic oscillations of $\xi^2_{\rm opt}$ and $\zeta^2_{\rm opt}$. 
Indeed, a non-zero value of the non-diagonal terms of the covariance matrix $\gamma_{yz}$ varying in time implies that $\xi^2_{\rm opt}$ and $\zeta^2_{\rm opt}$ 
are minimized along a direction in the $y$-$z$ plane that varies during the evolution.
For values of $\Lambda$ sufficiently far from the critical value $\Lambda = N/(N+1) \approx 1$, our Gaussian approximation (\ref{GState}) reproduces numerical results well.
We find a periodic generation of entanglement \cite{LawPRA2001}, see Fig.~\ref{figpista}, with minima at $2 \omega_\pi t = n \pi$, with $n$ an integer number,
reaching a minimum value (in time) of
\be \label{xi_stable}
\zeta^2_{\rm min} = 1 - \Lambda.
\ee
As expected, due to the relative weak nonlinearity, we obtain a modest generation of entanglement.
Approaching $\Lambda=1$ the Gaussian approximation reproduces the numerical findings only for very short times $t \ll \pi/\omega_\pi$, 
where $\xi^2_{\rm opt} \simeq \zeta^2_{\rm opt}$, as shown in the lower panel of Fig.~\ref{figpista}. 
For longer times, anharmonic effects in the potential $V(\phi)$ become important and cause damping of the
oscillations of $\xi^2_{\rm opt}$ and $\zeta^2_{\rm opt}$ that is not captured by the Gaussian approximation.

\begin{figure}[t!]
\includegraphics[width=0.46\textwidth]{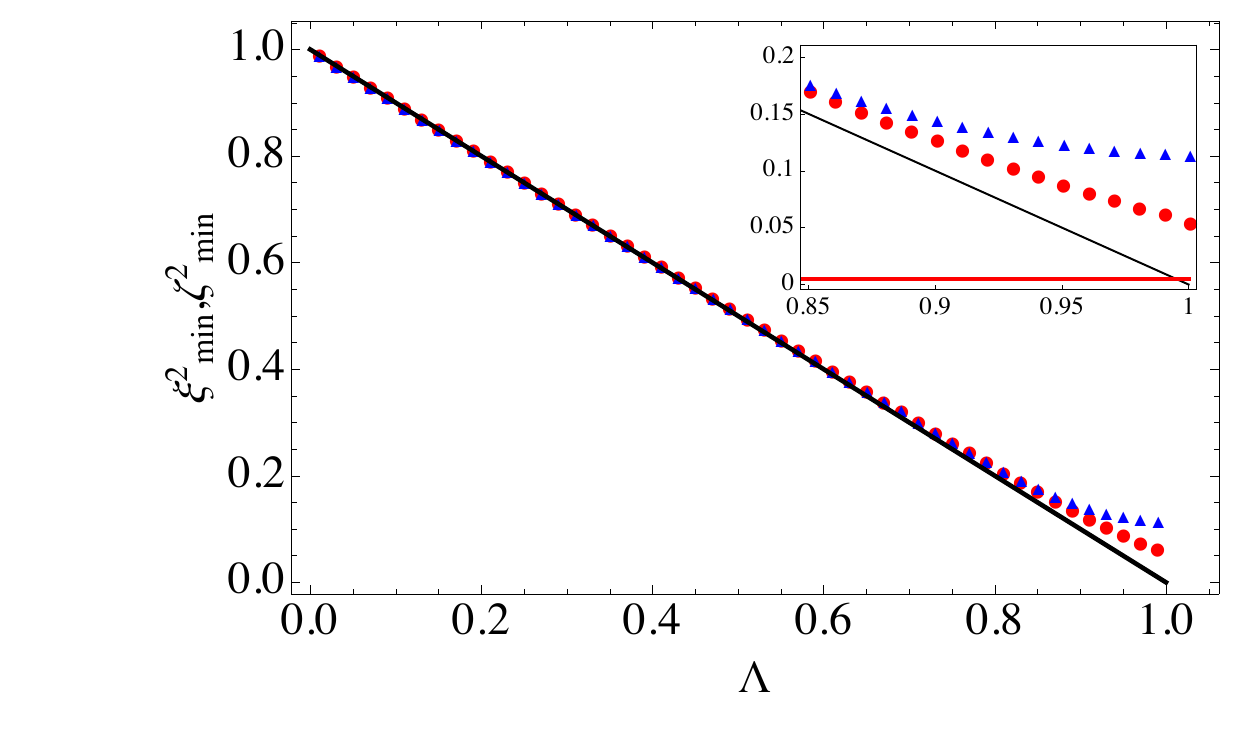}
\caption{(color online) Comparison between the minima (in time) of $\zeta^2_{\rm opt}$ (red dots) and $\xi^2_{\rm opt}$ (blue triangles) 
computed numerically and the analytical prediction Eq.~\rref{xi_stable} (black line).
The inset is a zoom around $\Lambda=1$. The horizontal red line is the Heisenberg limit $\zeta^2 = 1/N$. Here $N=200$.} 
\label{FigMin}
\end{figure}

In Fig.~\ref{FigMin} we plot the minimum values in time of $\xi^2_{\rm opt}$ and $\zeta^2_{\rm opt}$ as a function of $\Lambda$. The solid line is Eq.~(\ref{xi_stable}), while symbols are numerical results. 
We see that the maximal generation of entanglement depends on $\Lambda$ and Eq.~(\ref{xi_stable}), predicting an unphysical 
$\zeta^2_{\rm min} \to 0$ for $\Lambda \to 1$ and finite $N$ (which corresponds to a diverging QFI), fails when approaching the critical value.
The region where our approximation predicts the correct minima increases with the number of particles.

{\it Unstable regime $\Lambda > N/(N+1)$.}
In the unstable regime, the quantum dynamics of the initial CSS (\ref{CSS2}) is quite different from the stable one.
As shown in the right column of Fig.~(\ref{HusPi}) for $\Lambda=2$, the initial CSS symmetric Wigner distribution evolves quickly in a highly asymmetric ellipse, 
stretching along the separatrix of the mean-field dynamics.
This highly squeezed distribution leads us to expect a stronger generation of metrologically useful entanglement with respect to the stable regime.
For long times the state wraps around the mean-field fixed points at $\phi=\pi$ and $2\mean{\hat{J}_z}/N = \pm \sqrt{1+1/\Lambda^2}$ and 
large negative values of the Wigner function are obtained. 
In the unstable regime, the phase potential $V(\phi)$ is non-confining and the harmonic approximation is appropriate for relatively short times.
We find
\begin{subequations}
\label{expvalunpista}
 \begin{align}
 &\gamma_{zz} = \frac{1}{2(\Lambda-1)}\left(\Lambda+\Lambda \cosh 2{\omega}_\pi t -2 \right),\\
 &\gamma_{yy} = \frac{1}{2}\left(2-\Lambda+\Lambda \cosh 2{\omega}_\pi t \right),\\
 &\gamma_{yz}=\gamma{zy} = -\frac{2\Lambda}{\sqrt{1-\Lambda}}\sinh 2{\omega}_\pi t,
\end{align}
\end{subequations}
and
\be \label{expvalunpistb}
\frac{\langle\hat{J}_x\rangle}{N/2} = -1 + \frac{1}{4N}\frac{\Lambda^2}{\Lambda-1}\left(\cosh 2{\omega}_\pi t-1 \right).
\ee
The exponential time dependence of the matrix elements of $\gamma_{ij}$ implies a fast generation of entanglement.
It should be noticed that Eqs.~(\ref{expvalunpista}) and (\ref{expvalunpistb}) agree with those presented in~\cite{JuliaDiazPRA2012} and obtained 
with a different method (in number space~\cite{ShchesnovichPRA2008}) involving approximations similar to ours.

\begin{figure}[t!]
\includegraphics[width=0.46\textwidth]{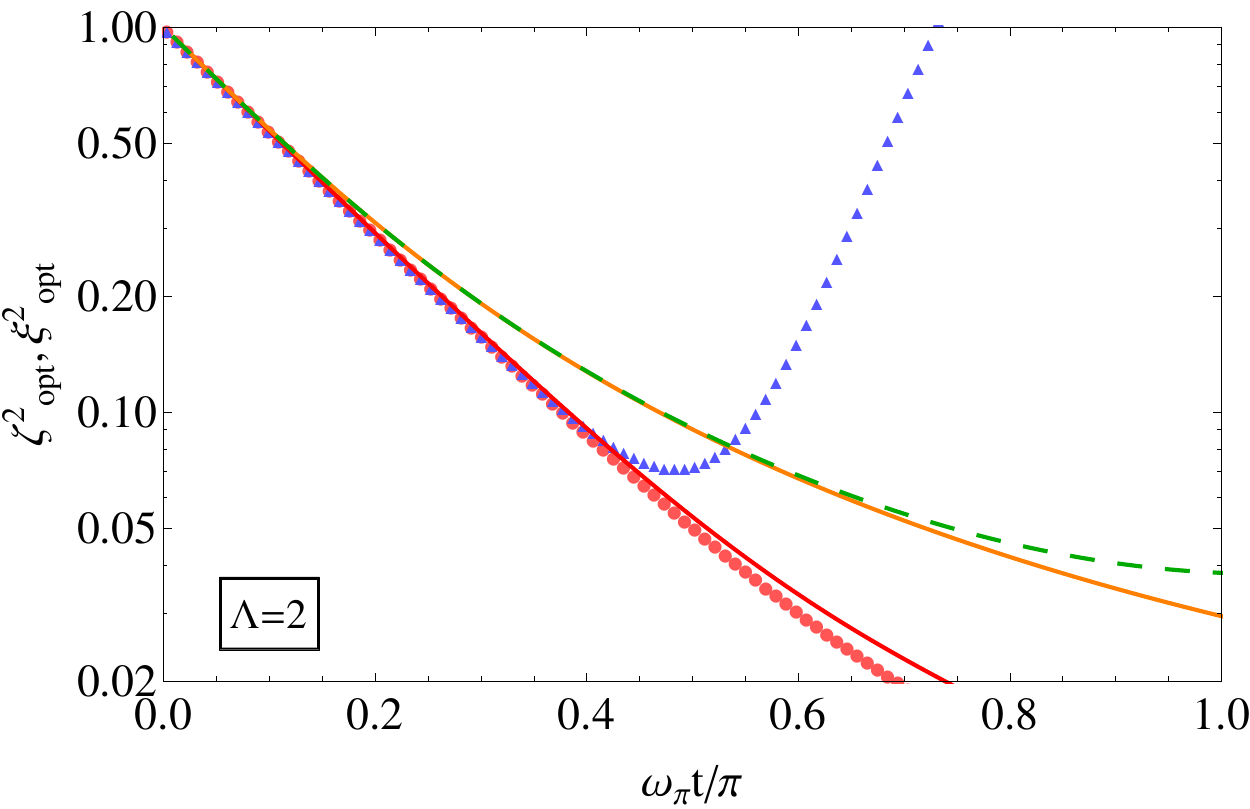}
\caption{(color online) 
Time evolution of $\zeta^2_{\rm opt}$ (red dots) and $\xi^2_{\rm opt}$ (blue triangles) compared to the analytical approximation for $\zeta^2_{\rm opt}$ (red line).
The green dashed line and the orange line are respectively $\xi^2_{\rm opt}$ and $\zeta^2_{\rm opt}$ for the one-axis twisting dynamics.
Here $N=200$.}
\label{ConfKU}
\end{figure}


In Fig.~\ref{ConfKU} we plot the analytical $\zeta^2_{\rm opt}$ (red line) compared with the results of QFI (red dots) and spin squeezing (blue triangles) obtained from 
the exact numerical solution of Eq.~(\ref{H}).
The spin-squeezing parameter decreases until it reaches a minimum and then it starts growing: the quantum dynamics features a loss of spin-squeezing 
at relatively long times ($\omega_{\pi} t \approx \pi/2$) that is associated to the wrapping of the state \cite{StrobelSCIENCE2014} and, equivalently,  
the onset of strongly negative parts of the Wigner function, as shown in Fig.~\ref{HusPi}.
In contrast, $\zeta^2_{\rm opt}$ keeps decreasing even after the spin-squeezing parameter reaches a minimum.
The QFI is a more powerful witness of entanglement in this case~\cite{StrobelSCIENCE2014}.
The analytical findings reproduce the short time dynamics quite well. 
The natural question is: how long does this agreement occur? 

\begin{figure}[tb]
\includegraphics[width=\columnwidth]{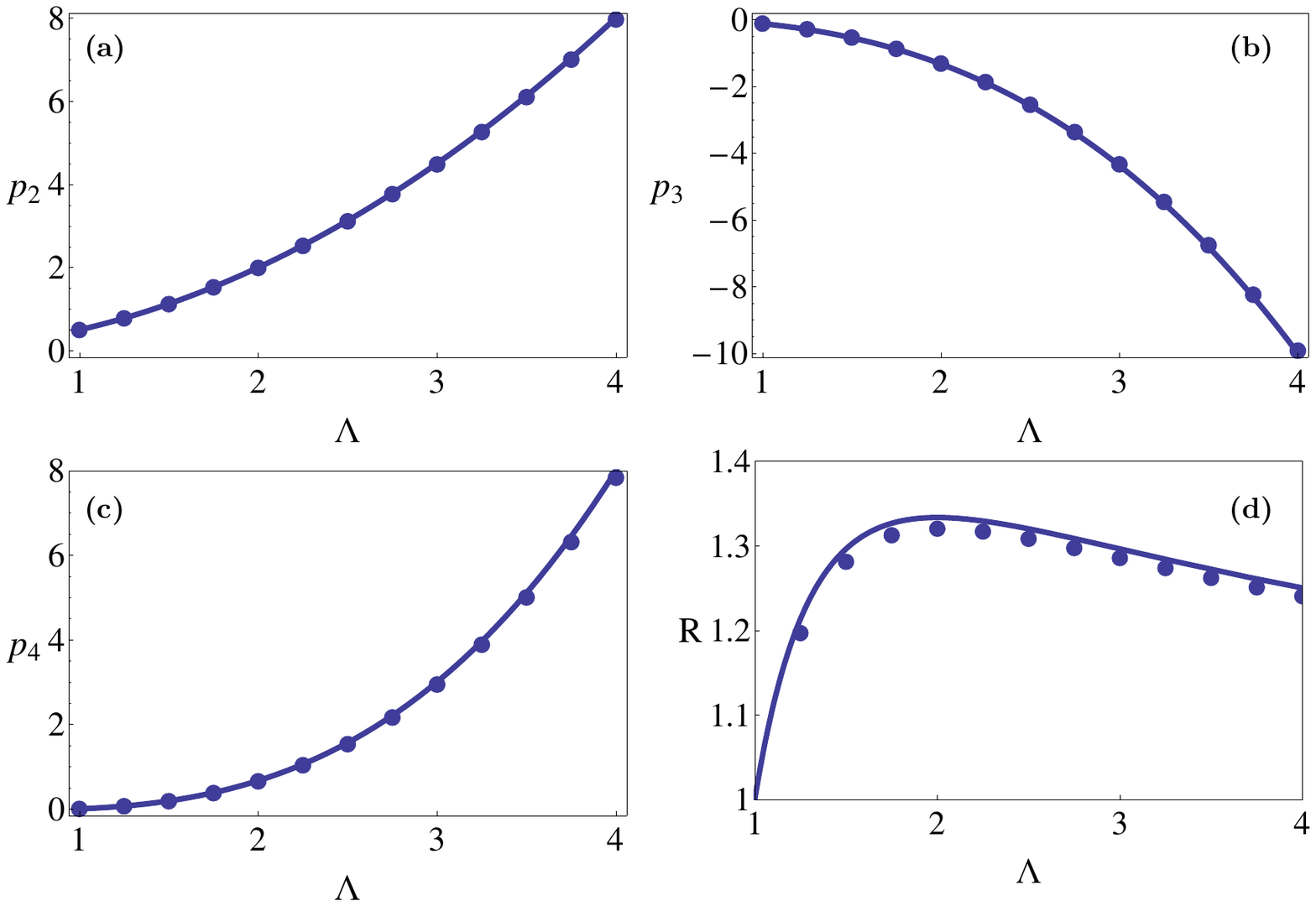}
\caption{
Panels (a), (b) and (c) show a comparison between the analytical (solid line) and the numerical (dots) values of $p_n$, with $n =1, 2, 3$ respectively, as a function of $\Lambda$. Here, $p_n$ is the coefficient of order $n$ in $N\chi t$ in the Taylor series of $\zeta^2$ divided by $\Omega^n$.
Lines are obtained from Eq.~(\ref{powunsta}): here $p_2 = \frac{\Lambda^2}{2}$, 
$p_3 = -\frac{\Lambda^3}{8} -\frac{\Lambda^2}{6} + \frac{\Lambda}{6}$ and $p_4 = \frac{1}{6}(\Lambda^3 - \Lambda^2)$.
Dots are obtained from a polynomial fit of $\zeta^2$ calculated with exact numerical diagonalization for $N=200$.
In panel (d) we show the ratio between the third order coefficient of the power series of $\zeta^2_{\pi}$ and of $\zeta^2_{\rm OAT}$.
Dots are numerical results and the line is Eq.~(\ref{EqR}).
}
\label{fit}
\end{figure}

Let us expand the $\zeta^2_{\pi} \equiv \zeta^2_{\rm opt}$ in Taylor series for short time
\begin{align} \label{powunsta}
\zeta^2_{\pi} &= 1 - N \chi t +\frac{N^2 \chi^2}{2} t^2 - \frac{N^3\chi^3}{8}t^3-\frac{1}{6}(\Lambda^2 - \Lambda)\Omega^3t^3 \notag\\&\quad+ \frac{1}{6}(\Lambda^3 - \Lambda^2)\Omega^4t^4+ \mathcal{O} (N\chi t)^5,
\end{align}
where $\Omega \lesssim N \chi$.
In Ref.~\cite{JuliaDiazPRA2012} it was argued that only the linear term in Eq.~(\ref{powunsta}) is accurate.
Figure~\ref{fit} shows that retaining only the linear term is overly conservative. 
There, we plot the coefficients of the Taylor expansion~(\ref{powunsta}) in comparison with those of a polynomial fit 
of the exact numerical result for Eq.~(\ref{H}). 
We find an extremely good agreement, even for relatively large values of $\Lambda$, up, at least, to the fourth order in $N \chi t$. 
The disagreement between the numerics and the analytical coefficients of the Taylor expansion is of the order $1/N$ that, as discussed above, is the accuracy of our approximations.
As we will show below, retaining higher-order terms in the Taylor expansion allows us to uncover the difference 
between the unstable dynamics and the one-axis twisting dynamics (i.e. $\Omega=0$). The two processes coincide up to linear~\cite{JuliaDiazPRA2012} and quadratic order, whereas the faster entanglement generation of the unstable dynamics can be observed in terms of third and fourth order.
  
\section{Comparison with the one-axis twisting dynamics}
\label{secOAT}

For the one-axis twisting evolution of an initial coherent spin state on the equator of the Bloch sphere, $\ket{\pi/2, \varphi}_N$ exact analytical results are available~\cite{KitagawaPRA1993}.
In particular
\be
\frac{\langle \hat{J}_x \rangle}{N/2} = \cos^{N-1}\left(\frac{\chi t}{2}\right), 
\ee
and
\be
 \lambda_\pm = 1 +\frac{1}{4}(N-1)\left[A \pm \sqrt{A^2+B^2}\right],
\ee
where $A= 1-\cos^{N-2}\left(\chi t\right)$ and $B = 4\sin\left(\frac{\chi t}{2}\right)\cos^{N-2}\left(\frac{\chi t}{2}\right)$.
From these equations we can readily obtain the optimized QFI $\zeta_{{\rm OAT}}^2$, using Eq.~(\ref{spinsq}).
Expanding in Taylor series, we obtain 
\be \label{powKU}
\zeta^2_{{\rm OAT}} = 1 - N \chi t +\frac{N^2 \chi^2 t^2}{2} - \frac{N^3\chi^3t^3}{8} + O(N\chi t)^5.\\
\ee
A direct comparison between Eq.~(\ref{powunsta})  and Eq.~(\ref{powKU}) shows that unstable dynamics and one-axis twisting produce, 
up to second order in $N \chi t$, the same amount of useful entanglement.
The leading difference appears in the third-order term:
\be 
\zeta^2_{{\rm OAT}} - \zeta^2_\pi = \frac{1}{6}\left(\Lambda^2 - \Lambda \right)\Omega^3t^3 + O(N\chi t)^4,
\ee
which is larger than zero for all $\Lambda >1$.
We also notice that the fourth-order term in $N \chi t$ is missing in Eq.~(\ref{powKU}), while it is present in 
Eq.~(\ref{powunsta}) and, being negative, it contributes to the generation of entanglement.
We thus conclude that for relatively short times (up to $O(N \chi t)^5$, at least), the unstable regime of the BJJ dynamics produce more entanglement than one-axis twisting. 
This provides an analytical explanation of the recent experimental findings of Ref.~\cite{MusselPRA2015, StrobelSCIENCE2014}.
The value of $\Lambda$ for which the generation of entanglement is the fastest can be obtained  
considering the ratio between the third-order coefficients in the power series of $\zeta_{\pi}$ and $\zeta_{{\rm OAT}}$:
\be \label{EqR}
R = 1+\frac{4}{3}\left(\frac{1}{\Lambda} - \frac{1}{\Lambda^2} \right).
\ee
The maximization of this equation provides $\Lambda = 2$.
In Fig.~\ref{fit} we compare our analytical predictions with exact numerical simulations.
We obtain a very good agreement that confirms $\Lambda=2$ as the optimal parameter for the quantum dynamics.
It should be noticed that $\Lambda=2$ is a special value in the mean-field BJJ dynamics: 
for $\Lambda=2$ the separatrix passing through the point $\phi=\pi$ and $z = 2 \mean{\hat{J}_z}/N =0$ reaches its maximum extension 
passing the poles of the Bloch sphere, $\phi=\pi$ and $z = 2 \mean{\hat{J}_z}/N =\pm 1$.
In a sense, the mean-field bifurcation attains its maximum criticality. 
For $\Lambda > 2$ the mean-field dynamics features the onset of macroscopic selftrapping  \cite{SmerziPRL1997,RaghavanPRA1999}.
Interestingly, $\Lambda=2$ has been also identified in Ref.~\cite{MicheliPRA2003} as the optimal condition for the dynamical creation 
of maximally entangled states [on time scales $\chi t \approx \log(8N)/N$ that are much longer than those considered here].

Finally, in Fig.~\ref{ConfKU} we compare $\zeta^2_{\rm opt}$  and $\xi^2_{\rm opt}$ for the unstable fixed-point dynamics
with the corresponding values for the OAT. 
The two different dynamics are equivalent for very short times. 
Yet, for sufficiently long times, when higher-order terms in the Taylor expansion~(\ref{powunsta}) become relevant, 
the generation of entanglement in the bosonic Josephson junction is faster than in the OAT model.
We can also notice that in both models, for times longer than those discussed in this work, there is a generation of states
which are not spin-squeezed but are recognized as entangled by the QFI.



\section{Entanglement generation around \texorpdfstring{$\mbox{\boldmath $\phi=0$}$}{}}
\label{Sta}

For completeness and comparison, here we study the quantum dynamics for a CSS
\be \label{CSS1}
\ket{\pi/2, 0}_N = \bigg( \frac{\hat{a}^\dagger + \hat{b}^\dagger}{\sqrt{2}} \bigg)^N\ket{{\rm vac}}, 
\ee
polarized along the positive $x$-axis (i.e., the eigenstate of $\hat J_x$ with maximum eigenvalue $N/2$).
An analytical study of the quantum dynamics close to the point $\phi \approx 0$ can be done substituting Eq.~\rref{EQPM} with the Hamiltonian of a harmonic oscillator of square frequency 
\be
\label{omega_0}
\omega_0^2 = 1 + \Lambda \Big(1+\frac{1}{N} \Big).
\ee
The frequency $\omega_0^2$ is always positive testifying that close to $\phi \approx 0$ we have an upward parabolic potential for every value of $\Lambda$, see Fig. \ref{figHarm}. 

\begin{figure}[t!]
  \includegraphics[width=0.45\textwidth]{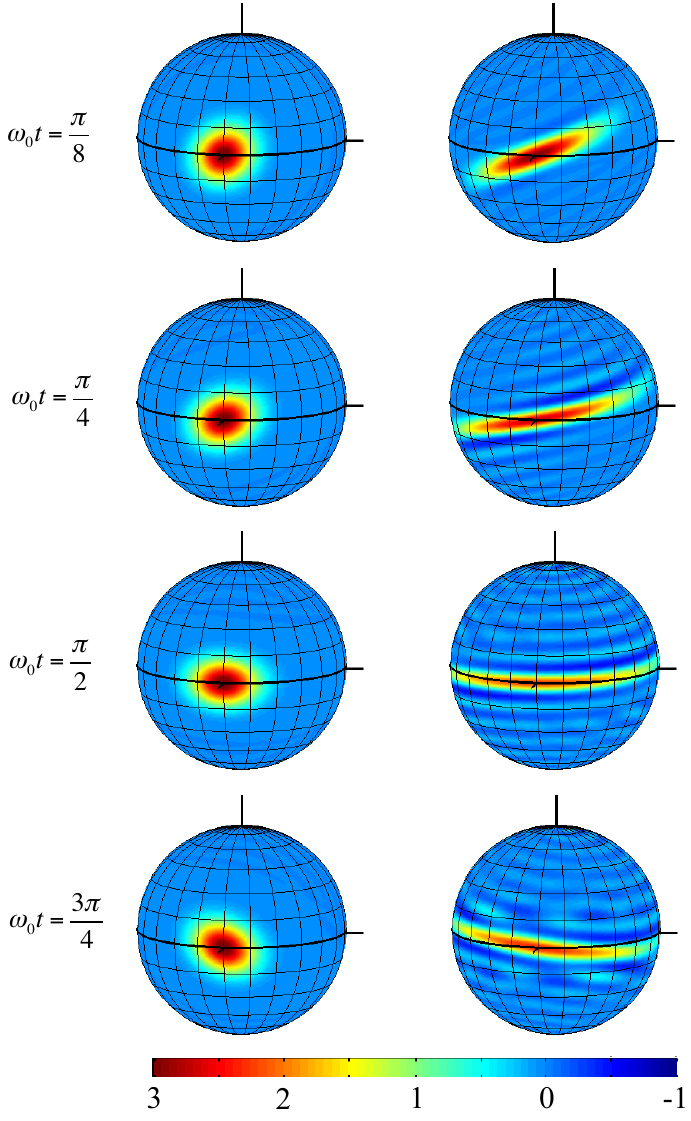}
 \caption{(color online) Wigner distributions (normalized to one) of the states obtained from the dynamical evolution of Eq.~(\ref{CSS1}) in the case $\Lambda=0.5$ (left column) and 
 $\Lambda=N$ (right column). Different rows correspond to different evolution times. Here $N=30$.}
 \label{Hus0}
\end{figure}

\begin{figure}[t!]
\includegraphics[width=0.46\textwidth]{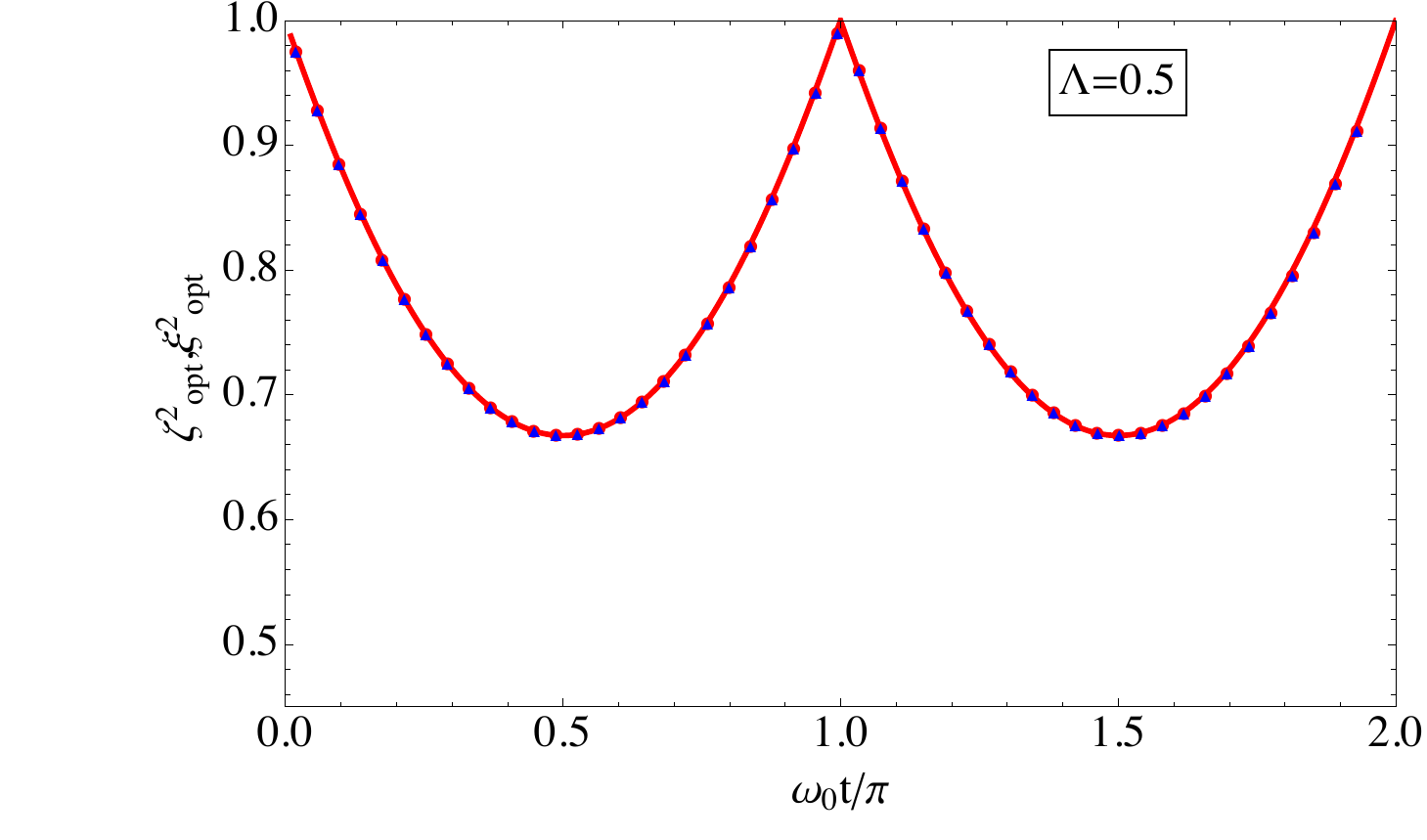}
\includegraphics[width= 0.46\textwidth]{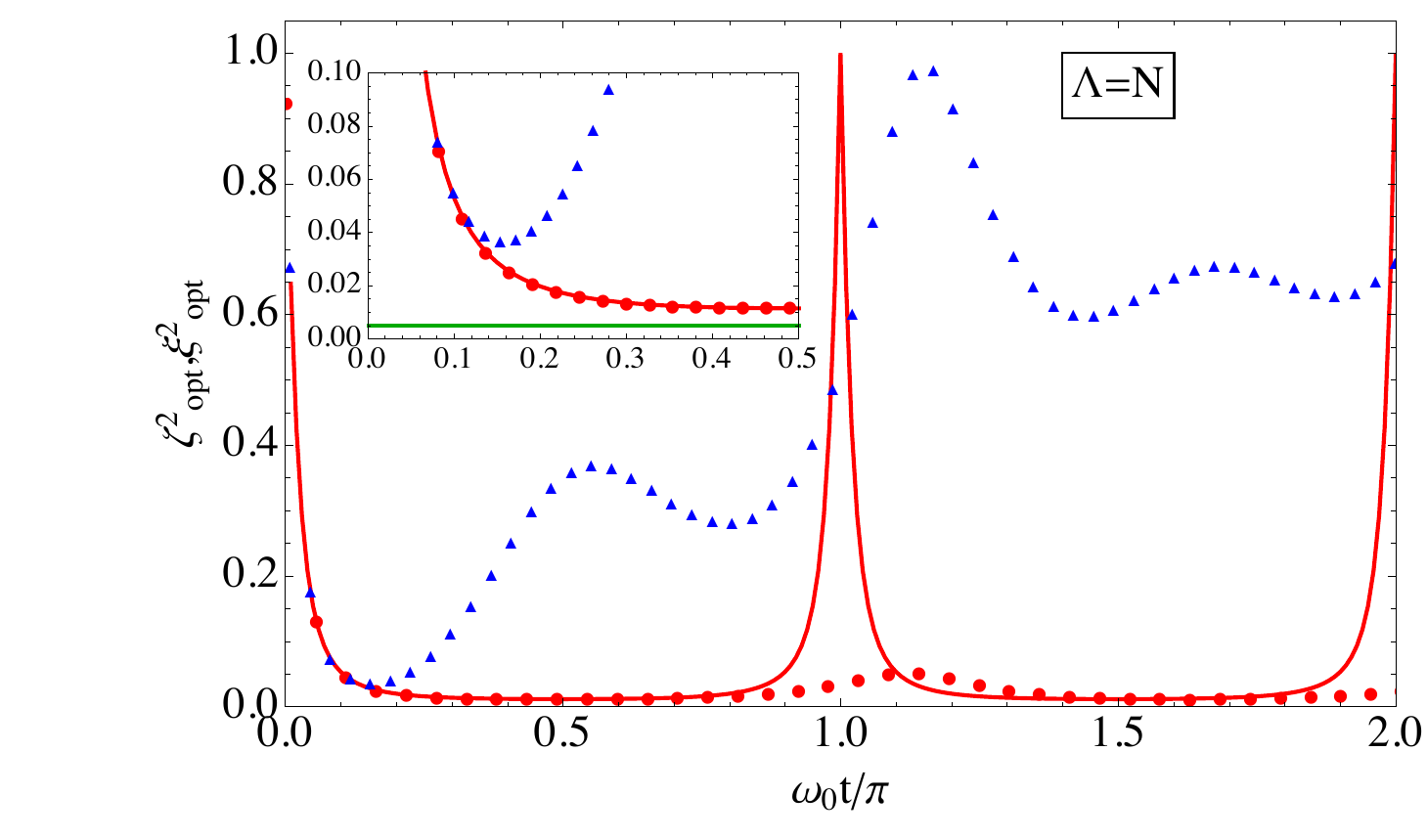}
\caption{(color online) 
Exact numerical time evolutions of $\zeta^2_{\rm opt}$ (red dots) and $\xi^2_{\rm opt}$ (blue triangles) along the direction of maximal entanglement.
The solid red line is the analytical result for $\zeta^2_{\rm opt}$. 
The two panels are obtained for different values of $\Lambda$ for $N=200$ and the initial state Eq.~\rref{CSS1}. The inset in the lower panel is a zoom of the short time behaviour, the horizontal green line indicates the Heisenberg limit.}
\label{figRabi}
\end{figure}

In Fig.~\ref{Hus0} we show the Wigner distribution plotted at different evolution time and $\Lambda$.
For $\Lambda \lesssim 1$, the dynamics is periodic and quite similar to the one observed in Fig.~\ref{HusPi} for $\phi=\pi$ at same value of $\Lambda$.
The main difference here is the occurrence of squeezing of the initially isotropic distribution along the $z$-axis.
Comparing the two columns we can see that increasing the value of $\Lambda$ also the asymmetry of the Wigner distribution increases
and negative parts appear. 
We study the quantum dynamics using the time-dependent wave function
\be
\Psi(\phi,t) = e^{-[a_0(t)+ib_0(t)]\phi^2},
\ee
with parameters
\begin{subequations} 
\label{ab}
\begin{align}
&a_0(t) = \frac{\frac{N\omega_0^2}{2\Lambda}\alpha_0}{\frac{\omega_0^2}{\Lambda^2}+\alpha_0^2 + \left(\frac{\omega_0^2}{\Lambda^2}-\alpha_0^2 \right)\cos (2\omega_0 t )} +\frac{N}{4\Lambda}\\
&b_0(t) = \frac{N\omega_0}{4\Lambda}\frac{\left(\frac{\omega_0^2}{\Lambda^2}-\alpha_0^2\right)\sin(2\omega_0 t)}{\frac{\omega_0^2}{\Lambda^2}+\alpha_0^2 + \left(\frac{\omega_0^2}{\Lambda^2}-\alpha_0^2\right)\cos(2\omega_0 t)},
\end{align}
\end{subequations}
where $\alpha_0 = \big(\frac{1+\omega_0(\Lambda_0)}{\Lambda_0}-\frac{1}{\Lambda}\big)$.
%
We obtain
\begin{subequations}
\label{expval0}
 \begin{align}
 &\gamma_{zz} = \frac{1}{2(1+\Lambda)}\left(2+\Lambda+\Lambda \cos 2\omega_0 t \right),\\
 &\gamma_{yy} = \frac{1}{2}\left(2+\Lambda-\Lambda \cos 2\omega_0 t \right),\\
  &\gamma_{yz}=\gamma_{zy} = \frac{\Lambda}{2\sqrt{1+\Lambda}}\sin2\omega_0 t.
\end{align}
\end{subequations}
giving the elements of the covariance matrix (\ref{cov}), and 
\be
\frac{\langle\hat{J}_x\rangle}{N/2} = 1 + \frac{1}{4N}\frac{\Lambda^2}{1+\Lambda}\left(\cos 2\omega_0 t-1 \right).
\ee
These expressions agree with those reported in \cite{JuliaDiazPRA2012}. 
As in the stable regime, nearby $\phi=\pi$ we have non-zero values of the non-diagonal terms of the covariance matrix $\gamma_{yz} = \gamma_{zy}$
that varies in time telling us that $\xi^2_{\rm opt}$ and 
$\zeta^2_{\rm opt}$ are minimized along a direction in the $y$-$z$ plane that varies during the evolution.
In Fig. \ref{figRabi} we show the oscillation of $\zeta^2_{\rm opt}$ for different values of $\Lambda$.
In the regime $\Lambda \lesssim 1$, where $\zeta^2_{\rm opt}$ and $\xi^2_{\rm opt}$ coincide, our approximation matches the exact numerical points excellently. 

The minima (in time) $\zeta^2_{\rm opt}$ are obtained for $2 \omega_0 t = n \pi$, with $n$ an integer number.
The depth of these minima, at the zeroth order in $1/N$, is: 
\be
\label{min}
\xi^2_{\min} \simeq \zeta^2_{\min} \sim \frac{1}{1+\Lambda}. 
\ee

\begin{figure}[t!]
\includegraphics[width=0.46\textwidth]{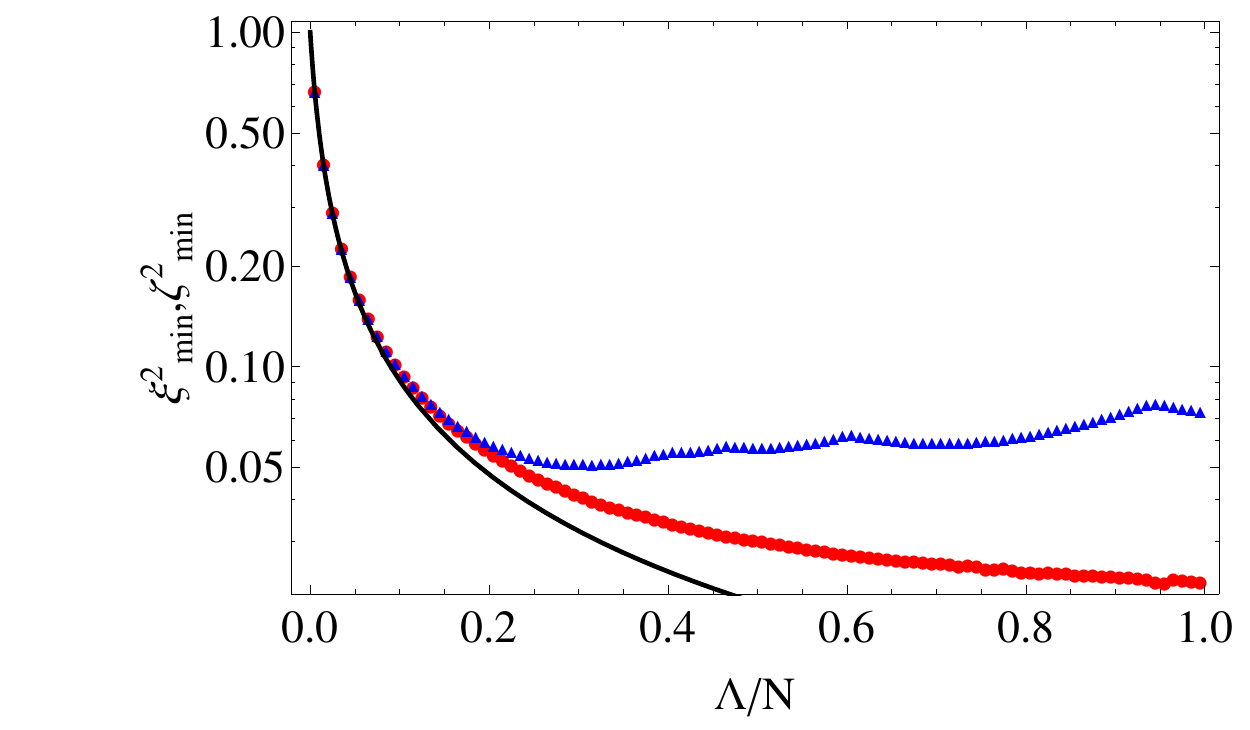}
\caption{(color online) Comparison between the minima (in time) of $\zeta^2_{{\rm opt}}$ (red dots) and $\xi^2_{{\rm opt}}$ (blue triangles)
computed numerically and the analytical prediction of Eq.~(\ref{min}) (black line).
}
\label{min0}
\end{figure}

According to this prediction, the entanglement increases with $\Lambda$.
However, the approximations leading to Eq.~(\ref{min}) fail for $\Lambda \gg 1$.
The lower panel of Fig.~\ref{figRabi} shows $\xi^2_{\rm opt}$ and $\zeta^2_{\rm opt}$ for $\Lambda =N$.
Our analytical results follow the dynamical regime only for relatively short times $t \lesssim \pi/\omega_0$.
For longer times we observe a net difference between the numerical $\xi^2_{\rm opt}$ and $\zeta^2_{\rm opt}$, 
corresponding to the onset of a negative Wigner function, see Fig.~\ref{Hus0}.
In Fig.~\ref{min0} we plot the minimum value (in time) of $\xi^2_{\rm opt}$ and $\zeta^2_{\rm opt}$ as a function of $\Lambda/N$.
We can see that the prediction of Eq. \rref{min} fails when $\Lambda$ becomes a significant fraction of the number of particles.

Finally, it is interesting to compare the stable and unstable dynamics for $\Lambda \gtrsim 1$. 
A Taylor expansion of $\zeta^2_{0} \equiv \zeta^2_{\rm opt}$ for $N\chi t \ll 1$ gives 
\begin{align}\label{powsta}
\zeta^2_0 &= 1 - N \chi t +\frac{N^2 \chi^2 t^2}{2} - \frac{N^3\chi^3t^3}{8}+\frac{1}{6}(\Lambda^2 + \Lambda)\Omega^3t^3 \notag\\&\quad + \frac{1}{6}(\Lambda^3 - \Lambda^2)\Omega^4t^4 + O(N \chi t)^5,  
\end{align}
where $\Omega \lesssim N \chi$.
A direct comparison with Eq.~(\ref{powunsta}) shows that all the terms in the Taylor expansion 
are equal except the sign of the $\Lambda^2$ factor in the third order term, which is positive in Eq.~(\ref{powsta})
and negative in Eq.~(\ref{powunsta}). For sufficiently short times, $\zeta^2_0 \leq \zeta^2_\pi$ for any value of $\Lambda$. 
Hence, the unstable regime at $\phi=\pi$ leads to a faster generation of entanglement on short time scales.

\section{Comparison to optimally controlled evolution}
Our results so far demonstrate that the unstable evolution at $\phi=\pi$ speeds-up the generation of entanglement in a bosonic Josephson junction 
when compared to the one-axis-twisting evolution ($\Omega=0$).
This effect is strongest at the optimal value $\Lambda=2$. 
In this section we compute an optimal control of the evolution under the  BJJ Hamiltonian.
\begin{align}\label{Hopt}
\hat{H}=\hbar\chi c \hat{J}_{\mathbf{n}}^2 - \hbar\Omega\hat{J}_{\mathbf{m}}
\end{align}
with adjustable parameters $-1\leq c\leq 1$, $\Omega\geq 0$, and normalized $\mathbf{n},\mathbf{m}\in\mathbb{R}^3$, 
in order to have the fastest possible increase of the quantum Fisher information. 
We consider the initial state $|\psi(0)\rangle=\ket{\pi/2, \pi}_N$, introduced in Eq.~(\ref{CSS2}). To identify a step-wise optimized dynamics, we discretize the time evolution into steps of length $\delta \tau$, and the state at $t=k \delta \tau$ is obtained as
\begin{align}\label{optimize}
|\psi(k \delta \tau)\rangle &= e^{-i\hat{H}_k\delta\tau}|\psi((k-1)\delta \tau)\rangle,
\end{align}
where the $\hat{H}_k$ is chosen from the family Eq.~(\ref{Hopt}) with parameters such that $\zeta^2_{\rm opt}$ yields the minimal possible value for the state $|\psi(k \delta \tau)\rangle$.

\begin{figure}[tb]
\includegraphics[width=0.46\textwidth]{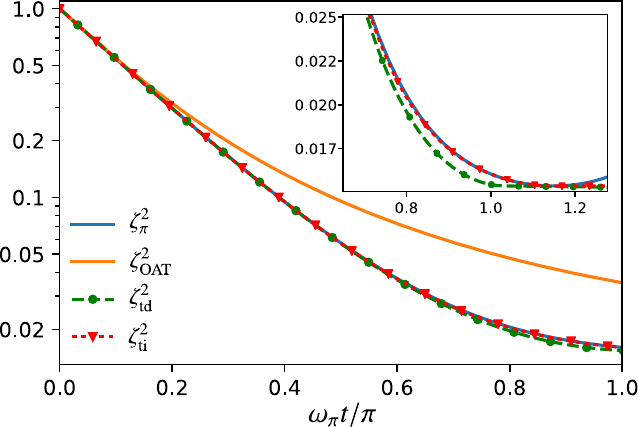}
\caption{(color online) Comparison of optimally controlled BJJ evolutions that maximize $\zeta^2_{\rm opt}$ with the unstable dynamics $\zeta^2_{\pi}$ and the one-axis twisting evolution $\zeta^2_{\rm OAT}$. For the optimal time-independent Hamiltonian, we propagate from $|\psi(0)\rangle$ to $|\psi(t)\rangle$ in a single step~(\ref{optimize}) with an optimally chosen Hamiltonian from the family~(\ref{Hopt}), producing $\zeta^2_{\rm ti}$. The optimally controlled evolution with a time-dependent Hamiltonian is obtained by discretizing the evolution time into small intervals and optimizing the Hamiltonian parameters before each consecutive timestep, leading to $\zeta^2_{\rm td}$. At short times, the optimal control schemes are unable to produce more entanglement than the evolution from the unstable fixed point. At longer times, the time-dependent scheme shows small enhancements (inset).}
\label{figOptControl}
\end{figure} 

The optimal time-independent Hamiltonian $\hat{H}$ for the generation of maximal amounts of useful entanglement after a time $t$ is obtained by performing the maximization~(\ref{optimize}) in a single step of length $\delta\tau = t$. Conversely, to allow for arbitrary time-dependent control schemes, we let $\delta\tau\rightarrow 0$. Fig.~\ref{figOptControl} compares the evolution of $\zeta^2_{\rm opt}$ for the numerically obtained optimal Hamiltonian (both time-dependent $\zeta^2_{\rm td}$ and time-independent $\zeta^2_{\rm ti}$) to the unstable dynamics [$c=1$, $\Omega=\chi N/2$, $\mathbf{n}=\mathbf{z}$, and $\mathbf{m}=-\mathbf{x}$] and the one-axis-twisting evolution [$c=1$, $\Omega=0$, $\mathbf{n}=\mathbf{z}$]. These results confirm the optimality of the unstable evolution for fast entanglement generation at short times, as neither of the two optimal-control schemes are able to produce larger amounts of entanglement at short times. As we approach the time where $\zeta^2_{\pi}$ attains its minimal value, very small enhancements in the entanglement production are achieved by the optimally controlled evolution with a time-dependent Hamiltonian (see inset). This evolution involves heavily oscillating Hamiltonian parameters \cite{footnote1}, which renders an experimental implementation improbable.


\section{Summary and conclusions}
To summarize, we have studied the generation of useful entanglement for quantum metrology in the quantum dynamics of a bosonic Josephson junction.
The dynamics is characterized by particle-particle interaction and linear coupling.
We have discussed different dynamical regimes, depending of the initial polarization of the coherent spin state and the value of interaction-over-tunneling parameter $\Lambda$.
The different regimes are well understood within a mean-field approximation.  
Analytical results for the quantum Fisher information, valid for relatively short times, lead us to two important results:
i) the dynamical generation of entanglement is fastest when the coherent spin state points along the negative $x$-axis and $\Lambda=2$, 
corresponding to maximum criticality in the mean-field unstable fixed point dynamics, and 
ii) in this regime, a direct comparison with the one-axis twisting dynamics \cite{KitagawaPRA1993, PezzePRL2009}
shows that linear coupling accelerates the dynamical creation of entanglement. 
Additional numerical analysis of optimal control schemes demonstrates the optimality of the evolution at 
$\Lambda=2$ for fast entanglement generation with bosonic Josephson junctions.
These findings help the understanding of recent experimental results \cite{MusselPRA2015} and 
can find direct application as a guideline to future experiments using an unstable dynamics to create large-scale entanglement.  

\section*{Acknowledgement}
M.G. acknowledges funding by the Alexander von Humboldt foundation.
This work has been supported by the European Commission through the QuantERA projects ``Q-Clocks" and  ``CEBBEC" and
the Empir project ``USOQS''.

\appendix 

\begin{widetext}
\section{Action of collective spin operators in the EQPM}
\label{App0}
In this Appendix we report how the actions of the spin operators on a general two-mode state can be expressed in terms of differential operators acting on $\Psi(\phi,t)$ defined in \rref{barg}. 
\begin{subequations}
\label{OP}
\begin{align}
&\hat{J}_x\ket{\psi(t)} =\int_{-\pi}^\pi  \frac{\mathrm{d}\phi}{2\pi}\ket{\phi}\left[\sin\phi\frac{\partial}{\partial\phi} +\left(\frac{N}{2}+1\right)\cos\phi\right]\Psi(\phi,t),\\
&\hat{J}_y\ket{\psi(t)} =\int_{-\pi}^\pi  \frac{\mathrm{d}\phi}{2\pi}\ket{\phi}\left[\cos\phi\frac{\partial}{\partial\phi} -\left(\frac{N}{2}+1\right)\sin\phi\right]\Psi(\phi,t),\\
&\hat{J}_z\ket{\psi(t)} = i\int_{-\pi}^\pi  \frac{\mathrm{d}\phi}{2\pi}\ket{\phi}\frac{\partial}{\partial\phi}\Psi(\phi,t).
\end{align}
\end{subequations}
From the above expressions we can obtain the action of second-order operators such that: 
\begin{subequations}
\label{OP2}
\begin{align}
\hat{J}^2_x\ket{\psi(t)} &= \int_{-\pi}^\pi  \frac{\mathrm{d}\phi}{2\pi}\ket{\phi} \left[ \sin^2\phi\frac{\partial^2}{\partial\phi^2} -\frac{N+3}{2}\sin (2\phi) \frac{\partial}{\partial\phi} + \left(\left(\frac{N}{2} +1\right)^2 - \left(\frac{N}{2} +2\right)\right)\cos^2\phi + \left(\frac{N}{2} -1\right) \right] \Psi(\phi,t)\\
\hat{J}^2_y\ket{\psi(t)} &= \int_{-\pi}^\pi  \frac{\mathrm{d}\phi}{2\pi}\ket{\phi} \left[ \cos^2\phi\frac{\partial^2}{\partial\phi^2} -\frac{N+3}{2}\sin (2\phi) \frac{\partial}{\partial\phi} + \left(\left(\frac{N}{2} +1\right)^2 + \left(\frac{N}{2} +1\right)\right)\sin^2\phi - \left(\frac{N}{2} +1\right) \right] \Psi(\phi,t),\\
\hat{J}^2_z\ket{\psi(t)}&=-\int_{-\pi}^\pi  \frac{\mathrm{d}\phi}{2\pi}\ket{\phi}\frac{\partial^2}{\partial\phi^2}\Psi(\phi,t),\\
\lbrace\hat{J}_y,\hat{J}_z\rbrace\ket{\psi(t)} &= i\int_{-\pi}^\pi  \frac{\mathrm{d}\phi}{2\pi} \ket{\phi}\left[2\cos \phi \frac{\partial^2}{\partial\phi^2} -(N+3)\sin \phi \frac{\partial}{\partial \phi} -\left( \frac{N}{2}+1\right)\cos \phi \right] \Psi(\phi,t). 
\end{align}
\end{subequations}

\section{Gaussian approximation}
\label{AppI}

The exact representation of the coherent states \rref{CSS2} and \rref{CSS1} in the EQPM is a Dirac delta, 
but for the calculation it is convenient to express everything in terms of Gaussian wave functions.
To do so we will not consider as initial state an eigenstate of the Hamiltonian \rref{H} with $\Lambda = 0$. 
We will instead initialize our EQPM dynamics with the Gaussian ground state of the harmonic approximation of \rref{EQPM} with an initial interaction parameter $\Lambda_0 \neq 0$
and then, at the end of the calculations, take the limit $\Lambda_0 \rightarrow 0$.

To compute $\xi^2_{\rm opt}$ and $\zeta^2_{\rm opt}$ in Section \ref{SSF} we need the expectation values of 
$\hat{J}^2_z$, $\hat{J}^2_y$, $\lbrace\hat{J}_z,\hat{J}_z\rbrace$ and $\hat{J}_x$. These operators can be expressed in the phase representation by equations \rref{OP}.
In this Appendix we present the procedure we have used to obtain analytical expressions for their expectation values.

Within our Gaussian approximation we have mapped the dynamics of the system into the well-known evolution of a Gaussian packet into a parabolic potential. Thus we have to deal with wave functions of the form 
\be
\Psi(\phi,t) = \mathcal{N} e^{-(a(t)+ib(t))\phi^2}.
\ee
Now we are going to compute the normalization constant $\mathcal{N}$:

\be
 \mathcal{N}^{-1} =  \int_{-\pi}^{\pi} \int_{-\pi}^{\pi} \Psi(\theta,t)^*\Psi(\phi,t) \braket{\theta}{\phi} d\theta d\phi 
 =\int_{-\pi}^{\pi} \int_{-\pi}^{\pi} e^{-(a(t)-ib(t))\theta^2} e^{-(a(t)+ib(t))\phi^2} \cos^N\left(\frac{\theta -\phi}{2}\right) d\theta d\phi.
\ee
This integral it is not easy to evaluate, but it can be approximated as a Gaussian integral that can be easily calculated.
This can be done noticing that the second-order power series of $\cos^N$ is equal to the one of a Gaussian of width $2/\sqrt{N}$,
\begin{subequations}
\begin{align}
\cos^N\left(\frac{\theta -\phi}{2}\right) &= 1 - \frac{N(\theta -\phi)^2}{8} +\mathcal{O}[(\theta -\phi)^{4}]\\
\mathrm{exp} \left(-\frac{N(\theta -\phi)^2}{8}\right) &= 1 - \frac{N(\theta -\phi)^2}{8} + \mathcal{O}[(\theta -\phi)^{4}].
\end{align}
\end{subequations}
Solving the generalized Gaussian integral we find 
\be
\mathcal{N}^{-1} = \frac{2\pi}{\sqrt{4(a(t)^2 +b(t)^2) + N a(t)}}.
\ee

We have seen that, given an operator $\hat{A}$ acting on a quantum state of the many-particle system $\ket{\psi}$, we can always reduce it to a differential operator $\tilde{A}$ acting on the wave function $\psi(\phi,t)$.
Thus the expectation value of $\hat{A}$ is obtained as
\be
\label{EXP}
\langle\hat{A}\rangle = \int_{-\pi}^{\pi} \int_{-\pi}^{\pi} e^{-(a(t)-ib(t))\theta^2} \left[\tilde{A}e^{-(a(t)+ib(t))\phi^2}\right] \cos^N\left(\frac{\theta -\phi}{2}\right) d\theta d\phi.
\ee
With the same procedure used for the normalization we can express \rref{EXP} in terms of Gaussian integrals. 
Thus, with these considerations and the expressions \rref{OP}, we can find approximated analytical expressions for all the expectation values needed to compute $\xi$ and $\chi$: 
\begin{subequations}
 \begin{align}
  \langle \hat{J}_x \rangle &= \frac{N}{2}e^{-\frac{8a(t)+N}{8\mathcal{A}}}\left[\frac{(\mathcal{A} +a(t))\cos(b(t)/\mathcal{A})+b(t)\sin(b(t)/\mathcal{A})}{\mathcal{A}}\right];\\
  \langle \hat{J}_z^2 \rangle &= \frac{N(a(t)^2+b(t)^2)}{\mathcal{A}};\\
  \langle \hat{J}_y^2 \rangle &= \frac{N}{8\mathcal{A}^2}\left\lbrace (N+1)\mathcal{A}^2+a(t) N\mathcal{A}- e^{-\frac{N+8a(t)}{2\mathcal{A}}}\left[ \vphantom{b(t)^2} 4b(t)\left(\mathcal{A}N-\mathcal{B}\right)\sin(4b(t)/\mathcal{A})+ \right. \right. \\ \nonumber
  &\quad+ \left. \vphantom{e^{-\frac{N+8a(t)}{2\mathcal{A}}}} \left. \left[(4a(t)^2+4b(t)^2)(4a(t)+\mathcal{B}-N\mathcal{A})-a(t)N^2(\mathcal{A} +2a(t))+4Nb(t)^2\right]\cos(4b(t)/\mathcal{A}) \right] \right\rbrace ;\\ 
  \langle \lbrace\hat{J}_y,\hat{J}_z\rbrace \rangle &= \frac{N}{2} \frac{e^{-\frac{8a(t)+N}{8\mathcal{A}}}}{\mathcal{A}^2} \left [ -(4a(t)(a(t)^2+b(t)^2)+Nb(t)^2)\cos(b(t)/\mathcal{A}) + \right. \\ \nonumber 
  &\quad+\left. b(t)(N\mathcal{A} - \mathcal{B})\sin(b(t)/\mathcal{A})\right ] ;
 \end{align}
\end{subequations}
where we have defined
\begin{subequations}
\begin{align}
 \mathcal{A} = 4[a(t)^2+b(t)^2] +a(t)N, \\
 \mathcal{B} = 4[a(t)^2+b(t)^2] -a(t)N.
\end{align}
\end{subequations}
\end{widetext}


\begin{thebibliography}{100}

\bibitem{PezzeRMP}
L. Pezz\`e, A. Smerzi, M.K. Oberthaler, R. Schmied, and P. Treutlein, 
Quantum metrology with nonclassical states of atomic ensembles, 
Rev. Mod. Phys. {\bf 90}, 035005 (2018).

\bibitem{KitagawaPRA1993}
M. Kitagawa and M. Ueda, 
Squeezed spin states, 
Phys. Rev. A {\bf 47}, 5138 (1993).

\bibitem{SorensenNATURE2001}
A.S. S\o rensen, L.-M. Duan, J. I. Cirac, and P. Zoller, 
Many- particle entanglement with Bose-Einstein condensates, 
Nature {\bf 409}, 63 (2001).

\bibitem{GrossNATURE2010}
C. Gross, T. Zibold, E. Nicklas, J. Est\`eve, and M. K. Oberthaler, 
Nonlinear atom interferometer surpasses classical precision limit, 
Nature {\bf 464}, 1165 (2010).

\bibitem{RiedelNATURE2010}
M.F. Riedel, P. B\"ohi, Y. Li, T.W. H\"ansch, A. Sinatra, and P. Treutlein, 
Atom-chip-based generation of entanglement for quantum metrology, 
Nature {\bf 464}, 1170 (2010).

\bibitem{OckeloenPRL2013}
C.F. Ockeloen, R. Schmied, M. F. Riedel, and P. Treutlein, 
Quantum metrology with a scanning probe atom interferometer,
Phys. Rev. Lett. {\bf 111}, 143001 (2013).

\bibitem{MuesselPRL2014}
W. Muessel, H. Strobel, D. Linnemann, D. B. Hume, and M. K. Oberthaler, 
Scalable spin squeezing for quantum-enhanced magnetometry with Bose-Einstein condensates, 
Phys. Rev. Lett. {\bf 113}, 103004 (2013).

\bibitem{SchmiedSCIENCE2016}
R. Schmied, J.-D. Bancal, B. Allard, M. Fadel, V. Scarani, P. Treutlein, and N. Sangouard, 
Bell correlations in a Bose-Einstein condensate,
Science {\bf 352}, 441 (2016).


\bibitem{LerouxPRL2010}
Leroux, I. D., M. H. Schleier-Smith, and V. Vuleti\'c, 
Implementation of cavity squeezing of a collective atomic spin, 
Phys. Rev. Lett. {\bf 104}, 073602 (2010). 

\bibitem{TakeuchiPRL2005}
M. Takeuchi, S. Ichihara, T. Takano, M. Kumakura, T. Yabuzaki, and Y. Takahashi,
Spin squeezing via one-axis twisting with coherent light, 
Phys. Rev. Lett. {\bf 94}, 023003 (2005).

\bibitem{Schleier-SmithPRA2010}
M.H. Schleier-Smith, I. D. Leroux, and V. Vuleti\'c, 
Squeezing the collective spin of a dilute atomic ensemble by cavity feedback, 
Phys. Rev. A {\bf 81}, 021804 (2010).

\bibitem{MilburnPRA1997}
G.J. Milburn, J. Corney, E. M. Wright, and D. F. Walls, 
Quantum dynamics of an atomic Bose-Einstein condensate in a double-well potential, 
Phys. Rev. A {\bf 55}, 4318 (1997).

\bibitem{CiracPRA1998}
J.I. Cirac, M. Lewenstein, K. M\o lmer, and P. Zoller, 
Quantum superposition states of Bose-Einstein condensates,
Phys. Rev. A {\bf 57}, 1208 (1998).

\bibitem{SteelPRA1998}
M. J. Steel and M. J. Collett, 
Quantum state of two trapped Bose-Einstein condensates with a Josephson coupling,
Phys. Rev. A {\bf 57}, 2920 (1998).

\bibitem{AnanikianPRA2006}
D. Ananikian and T. Bergeman, 
Gross-Pitaevskii equation for Bose particles in a double-well potential: Two-mode models and beyond, 
Phys. Rev. A {\bf 73}, 013604 (2006).

\bibitem{GatiJPA2007}
R. Gati and M.K. Oberthaler, 
A bosonic Josephson junction, 
J. Phys. B: At. Mol. Opt. Phys. {\bf 40}, R61 (2007).

\bibitem{SmerziPRL1997}
A. Smerzi, S. Fantoni, S. Giovanazzi, and S.R. Shenoy, 
Quantum coherent atomic tunneling between two trapped Bose- Einstein condensates,
Phys. Rev. Lett. {\bf 79}, 4950 (1997).

\bibitem{RaghavanPRA1999}
S. Raghavan, A. Smerzi, S. Fantoni, and S.R. Shenoy, 
Coherent oscillations between two weakly coupled Bose-Einstein condensates: Josephson effects, $\pi$ oscillations, and macroscopic quantum self-trapping, 
Phys. Rev. A {\bf 59}, 620 (1999).

\bibitem{MicheliPRA2003}
A. Micheli, D. Jaksch, J. I. Cirac, and P. Zoller, 
Many-particle entanglement in two-component Bose-Einstein condensates, 
Phys. Rev. A {\bf 67}, 013607 (2003).

\bibitem{StrobelSCIENCE2014}
H. Strobel, W. Muessel, D. Linnemann, T. Zibold, D. B. Hume, L. Pezz\`e, A. Smerzi, and M. K. Oberthaler, 
Fisher information and entanglement of non-Gaussian spin states, 
Science {\bf 345}, 424 (2014).

\bibitem{JuliaDiazPRA2012b}
B. Juli\'a-D\'iaz, E. Torrontegui, J. Martorell, J. G. Muga, and A. Polls,
Fast generation of spin-squeezed states in bosonic Josephson junctions, 
Phys. Rev. A {\bf 86}, 063623 (2012);
A. Yuste, B. Julia-Diaz, E. Torrontegui ,J. G. Muga, J. Martorell, A. Polls, 
Shortcut to adiabaticity in internal Josephson junctions, 
Phys. Rev. A {\bf 88}, 043647 (2013).

\bibitem{LapertPRA2012}
M. Lapert, G. Ferrini, and D. Sugny, 
Optimal control of quantum superpositions in a bosonic Josephson junction, 
Phys. Rev. A {\bf 85}, 023611 (2012).

\bibitem{MusselPRA2015}
W. Muessel, H. Strobel, D. Linnemann, T. Zibold, B. Juli\'a-Diaz, and M. K. Oberthaler, 
Twist-and-turn spin squeezing in Bose-Einstein condensates, 
Phys. Rev. A {\bf 92}, 023603 (2015).

\bibitem{AnglinPRA2001}
J.R. Anglin, P. Drummond, and A. Smerzi, 
Exact quantum phase model for mesoscopic Josephson junctions, 
Phys. Rev. A 4353 {\bf 64}, 063605 (2001).

\bibitem{PezzePRA2005}
L. Pezz\`e, L. A. Collins, A. Smerzi, G. P. Berman, and A. R. Bishop,
Sub-shot-noise phase sensitivity with a Bose-Einstein condensate Mach-Zehnder interferometer, 
Phys. Rev. A {\bf 72}, 043612 (2005).

\bibitem{LawPRA2001}
C.K. Law, H. T. Ng, and P. T. Leung,
Coherent control of spin squeezing, 
Phys. Rev. A {\bf 63}, 055601 (2001).

\bibitem{JuliaDiazPRA2012}
B. Juli\'a-Diaz, B., T. Zibold, M.K. Oberthaler, M. Mel\'e-Messeguer, J. Martorell, and A. Polls, 
Dynamic generation of spin-squeezed states in bosonic Josephson junctions,
Phys. Rev. A {\bf 86}, 023615 (2012).

\bibitem{WallsBOOK}
D. Walls, and G. Milburn, {\it Quantum Optics} (Springer, New York, 1994).

\bibitem{ArecchiPRA1972}
F.T. Arecchi, E. Courtens, R. Gilmore, and H. Thomas, 
Atomic coherent states in quantum optics,
Phys. Rev. A {\bf 6}, 4368 2211 (1972).

\bibitem{ShchesnovichPRA2008}
V.S. Shchesnovich and M. Trippenbach, 
Fock-space WKB method for the boson Josephson model describing a Bose-Einstein condensate trapped in a double-well potential, 
Phys. Rev. A {\bf 78}, 5754 023611 (2008).

\bibitem{JavanainenPRA1999}
J. Javanainen, and M. Yu. Ivanov, 
Splitting a trap containing a Bose-Einstein condensate: Atom number fluctuations, 
Phys. Rev. A {\bf 60}, 2351 (1999).

\bibitem{HelstromBOOK}
C.W. Helstrom, 
{\it Quantum Detection and Estimation Theory} (Academic Press, New York, 1976).

\bibitem{BraunsteinPRL1994}
S. L. Braunstein and C. M. Caves, 
Statistical distance and the geometry of quantum states, 
Phys. Rev. Lett. {\bf 72}, 3439 (1994).

\bibitem{WinelandPRA1992}
D.J. Wineland, J. J. Bollinger, W. M. Itano, F. L. Moore, and D. J. Heinzen, 
Spin squeezing and reduced quantum noise in spectroscopy, 
Phys. Rev. A {\bf 46}, R6797 (1992);
D.J. Wineland, J. J. Bollinger, W. M. Itano, and D. J. Heinzen,
Squeezed atomic states and projection noise in spectroscopy,
Phys. Rev. A {\bf 50}, 67 (1994).

\bibitem{PezzePRL2009}
L. Pezz\`e and A. Smerzi, 
Entanglement, nonlinear dynamics, and the Heisenberg limit, 
Phys. Rev. Lett. {\bf 102}, 100401 (2009).

\bibitem{HyllusPRA2012}
P. Hyllus et al., 
Fisher information and multiparticle entanglement, 
Phys. Rev. A {\bf 85}, 022321 (2012); 
G. T\'oth, 
Multipartite entanglement and high-precision metrology, 
Phys. Rev. A {\bf 85}, 022322 (2012).

\bibitem{GiovannettiPRL2006}
V. Giovannetti, S. Lloyd and M. Maccone, 
Quantum Metrology, 
Phys. Rev. Lett. {\bf 96}, 010401 (2006).

\bibitem{DowlingPRA1994}
J.P. Dowling, G. S. Agarwal, and W. P. Schleich, 
Wigner distribution of a general angular-momentum state: Applications to a collection of two-level atoms, 
Phys. Rev. A {\bf 49}, 4101 (1994).

\bibitem{footnote1}
The underlying optimal evolution involves an external field with an intensity $\hbar\Omega$ that slowly ramps down from an initial value of $\hbar \chi N/2$ (corresponding to $\Lambda=2$) to zero, with a heavily oscillating orientation $\mathbf{m}$ in the $\mathbf{x}-\mathbf{z}$ plane as well as an interaction that strongly fluctuates between $c=1$ along $\mathbf{m}=(\mathbf{y}+\mathbf{z})/\sqrt{2}$ and $c=-1$ along $\mathbf{m}=(\mathbf{y}-\mathbf{z})/\sqrt{2}$.


\end{thebibliography}
 

\end{document}